\documentclass[prb,twocolumn,superscriptaddress,showpacs]{revtex4}
\usepackage{graphicx}
\usepackage{float}
\usepackage{dcolumn}
\usepackage{bm}
\usepackage{mathrsfs}
\usepackage{ulem}
\usepackage[per-mode = fraction]{siunitx}
\usepackage{hyperref}
\usepackage{xcolor}
\usepackage{soul}
\usepackage{amsmath}
\usepackage{amssymb}

\begin{document}

\title{Possible realization of a phononic tsunami in a wedge-shaped sample}
\author{Yuriy Yerin}
\affiliation{ Dipartimento di Fisica e Geologia, Universitá degli Studi di Perugia, Via Pascoli, 06123 Perugia, Italy}
\author{Andrey Varlamov}
\affiliation{CNR-SPIN, via del Fosso del Cavaliere 100, 00133, Rome, Italy}
\author{Claudia Fasolato}
\affiliation{ Dipartimento di Fisica e Geologia, Universitá degli Studi di Perugia, Via Pascoli, 06123 Perugia, Italy}
\author{Francesco Sacchetti}
\affiliation{ Dipartimento di Fisica e Geologia, Universitá degli Studi di Perugia, Via Pascoli, 06123 Perugia, Italy}
\affiliation{CNR Istituto Officina dei Materiali (IOM), I-06123 Perugia, Italy}
\author{Paolo Postorino}
\affiliation{Universit\`{a} di Roma Sapienza, Dipartimento di Fisica, I-00100 Roma, Italy}
\affiliation{CNR Istituto Officina dei Materiali (IOM), I-06123 Perugia, Italy}
\author{Caterina Petrillo}
\affiliation{ Dipartimento di Fisica e Geologia, Universitá degli Studi di Perugia, Via Pascoli, 06123 Perugia, Italy}
\date{\today }

\begin{abstract}
Exploiting the theory of solitons in a nonlinear elastic medium we predict a novel phenomenon called a phononic tsunami, which is characterized by the dramatic increase of the local amplitude of phonon modes. To elucidate the possible experimental detection of this phenomenon we propose to use a wedge-shaped sample in which a sharp edge serves for the emulation of the shoaling effect and such a local enhancement can be observed. Together with eigenfrequencies of transverse and longitudinal phonon modes of a system we find the characteristic dispersion relations that can be considered as a hallmark of a phononic tsunami. We justify our predictions by means of analytical calculations and numerical simulations showing a possible realization of this nonlinear effect in such a geometry. Our results provide the framework for the implementation of new kind experiments aimed at realizing and investigating a phononic tsunami phenomenon in relevant materials.
\end{abstract}

\maketitle

\section{Introduction}

A tsunami is a sporadic powerful water wave triggered by a shock external force (usually an earthquake) that can travel for thousands of kilometers from the disturbance across deep ocean. In the open ocean the tsunami does not have a significantly large amplitude, but its wavelength is extremely long. Due to these properties, it is difficult to detect a tsunami before it nears shore. However, on approaching the coast, when the effect of shoaling becomes crucial, a small amplitude open-ocean tsunami evolves into a large amplitude wave, with the bottom topography altering considerably its characteristics. Tsunami at intermediate depth is described by the dispersion relation ${\omega ^2} = gk\tanh \left( {kh} \right)$ that can be obtained from the Airy wave theory \cite{Kundu, Craik} (here  $k$ is the wavenumber, $h$ is the equilibrium water depth and $g$ is the gravitational acceleration constant).

A series of different approaches related to tsunami wave propagation have been discussed in literature. Most of them include nonlinearly dispersive water wave models based on the soliton theory, like the Korteweg de Vries, and Boussinesq equations. Later the Boussinesq-type model was reformulated by  Peregrine  for long waves in shallow waters of varying depth. However, it is worth noting that the consideration of the tsunami as a manifestation of soliton physics is the still debated and controversial topic. \cite{Constantinadrian1, Constantinadrian2, Glimsdal}

Since sound propagation and heat diffusion can both be described as mechanical vibrations transmitted through a crystal exploiting the phonon picture, the emergence of similar complex phenomena and related topological effects, like a phononic tsunami in the lattice, can be anticipated. From the theoretical point of view some progress in that direction has been achieved already with the aim to show the possible diversity and plenty of extraordinary nonlinear phenomena occurring with phonons.
It was shown by means of a diagram technique that the classical vibrational degrees of freedom of a solid, being sufficiently far from equilibrium, can be evolved into the phononic turbulence due to nonlinear interactions between long-wavelength modes \cite{Gantsevich}. Later on, it was found that in a cylindrical quantum wire embedded in another material acoustic phonon modes give rise another hydrodynamic-like nonlinear topological excitation known as a phonon vortex with nonzero angular momentum along the wire axis \cite{Nishiguchi}. Moreover, based on two different approaches, the homogeneous Fermi-Pasta-Ulam-Tsingou model and the nonlinear Schr\"odinger equation, the formation of phononic rogue waves in phononic lattices was predicted \cite{Kevrekidis1}. Along with this prediction different theoretical studies and subsequent experimental observations show that hydrodynamic phonon transport occurs in materials with a high Debye temperature and large anharmonicity.  Interestingly, that this hydrodynamic-like phenomenon can be described by the macroscopic transport equation similar to the Navier-Stokes equation \cite{Lee}.

Recent technological progress on newly developed phononic crystals and devices, combined with theoretical modeling, enabled control over material properties providing unique opportunities to manage and manipulate the phononic spectrum and other characteristics of these systems \cite{Maldovan, Fahrat}.  As a result, a large variety of experiments can be designed to verify the theoretical predictions of the above-mentioned effects in the field of nonlinear phononics \cite{Cavalleri}. 
A recent possible confirmation of the existence of novel nonlinear phenomena, namely phononic solitons, was obtained in traveling four-wave mixing experiments with incorporation of chirped input pulses into nonlinear phononic crystal \cite{Kurosu}. Undoubtedly, this experiment opens up the possibility of the observing not only phononic solitons but also another closely related effect, the phononic tsunami.

To this end, the goal of the present paper is to provide a theoretical background and a description of the phononic tsumani phenomenon for its possible detection in relevant materials. It is worth note that the study of strongly nonlinear elastic waves propagating in a wedge-shaped waveguide and known as wedge waves is longstanding challenge \cite{Maradudin1, Lagasse, Loss, Krylov, Zavorokhin, Sokolova}. After the theoretical prediction of wedge waves subsequent experiments confirmed presence of nonlinearities induced by the laser-based pump-probe excitations \cite{Adler, Jia, Lomonosov}.

\begin{figure}[ht]
\centering
\includegraphics[width=0.495\textwidth]{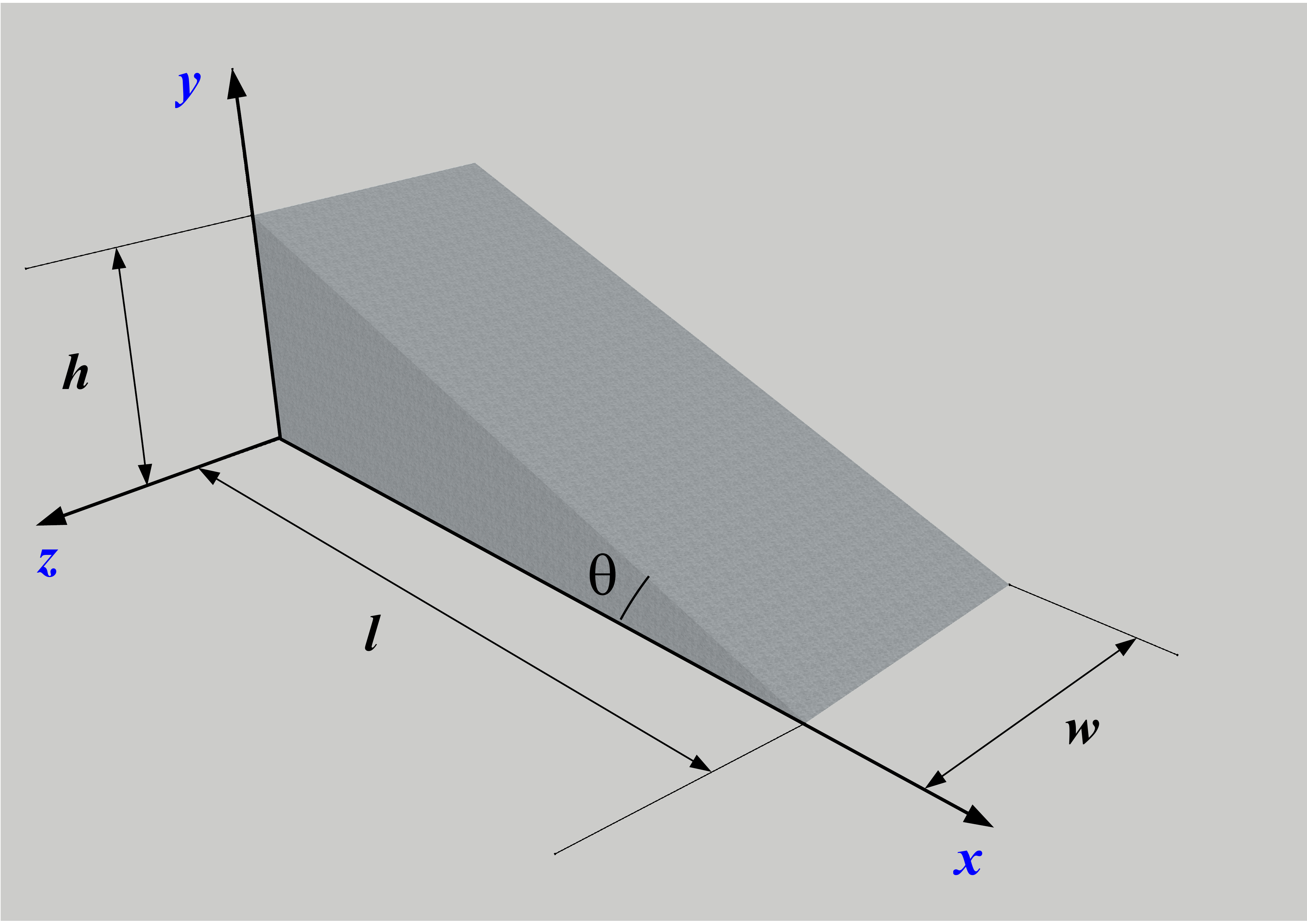}
\caption{Sketch of the sample geometry and the adopted reference system.}
\label{wedge}
\end{figure}

In turn, we propose to use a wedge-shaped sample to emulate the tsunami shoaling when by means of the shock wave pulse applied to a small side of the wedge, the phonon excitations approach its sharp edge (see Fig. \ref{wedge}). Since phonons can demonstrate hydrodynamic-like behaviour (turbulence, hydrodynamic transport, rogue waves) we define such a phenomenon in spirit of an ocean-wave tsunami. In other words, the phononic tsunami is characterized by the dramatic enhancement of the displacement field in a narrow edge of a wedge. It can be induced by the correspondingly applied shock wave in the specific geometry of the wedge-shaped sample. In this case the narrow edge of a wedge replicates the shoaling effect for the ocean tsunami wave, which runs from deep to shallow water. The energy of the tsunami wave is concentrated at the edge tip and as a result high displacement amplitude is achieved.  Exploiting the equation for modeling of solitons in a nonlinear elastic media, we find dispersion relations that can be considered as a hallmark of the phononic tsunami occurrence \cite{note0}. 

The paper is organized as follows. In sec. II we present the model and determine the eigenfrequencies for transverse and longitudinal phonon modes of a wedge-shaped sample. The equation describing propagation of the phononic tsunami is presented and discussed in  Sec. III, where we also find corresponding dispersion relations. Numerical simulations of the phenomenon based on the introduced equation for the wedge-shaped geometry are presented in Sec. IV. In Section V we study analytically the dynamical behavior of a phononic tsunami induced the excitation of the specific form of a short Gaussian pulse. The results obtained are summarized in Section VI.

\section{Model and Eigenfrequencies of phonon modes}

The system under consideration is modeled as a wedge with the length $l$, the height $h$, the width $w$ and the angle $\theta  = \arctan \frac{h}{l}$ (Fig. 1).  

Before the description of a phononic tsunami phenomenon the identification of the eigenfrequencies of all possible phonon modes for this geometry is required. One has to distinguish the longitudinal and transverse  modes and their eigenfrequencies corresponding to the different boundary conditions. This information is necessary for the possible experimental identification of a phononic tsunami by means of dispersion relations describing its propagation in a wedge-shaped sample (see Section III). From the formal point of view, we suggest using the same strategy as it is usually proposed for the ocean tsunami early warning system. 

We introduce the coordinate system as shown in Figure 1. To characterize transverse phonon modes in a wedge we define the displacement field as a vector ${\mathbf{u}}$ directed along the $z$-axis, yet independent on  $z$ (i.e. ${\mathbf{u}}$ has only one component $u_z = u(x, y)$). What concerns the longitudinal phonon modes the vector ${\mathbf{u}}$ of corresponding displacement field belongs to
 the $x$-$y$ plane. Without loss of generality one can assume that it has only single component $u_x=u(x,y)$. Finally, to simplify the model we consider it without defects.\cite{Kosevich_book, Mermin}.

\subsection{Eigenfrequencies of transverse phonon modes}

When the phonon wavelength is larger than the width of a wedge, the transverse phonon modes dominate and the biharmonic equation should be applied for their description \cite{Landau, Timoshenko} 
\begin{equation}
\label{biharmonic}
\rho \frac{{{\partial ^2}u}}{{\partial {t^2}}} =  - \frac{D}{w}{\Delta ^2}u,
\end{equation}
where the material dependent parameters $\rho$ and $D = \frac{{E{w^3}}}{{12\left( {1 - {\nu ^2}} \right)}}$  denote mass density and flexural rigidity of the material, $E$  is the Young's modulus,  $\nu$ is Poisson's ratio,  and ${\Delta ^2}u = {\nabla ^4}u$ is the Bi-laplacian operator. The displacement field for transversal phonon modes is calculated within the approximation that the width $w$ of a wedge is much smaller than its length and the height. Due to this assumption the consideration of transverse phonon modes for the three-dimensional geometry (see Fig. \ref{wedge}) is reduced to the quasi-two-dimensional system (triangle).

Corresponding boundary conditions depend on the fact whether the wedge edges are clamped (fixed) or free.  In the case of fixed edges the boundary conditions are
\begin{equation}
\label{bc_fixed}
\begin{gathered}
  {\left. u \right|_{x = 0,x = l}} = 0,{\text{ }}{\left. {\frac{{\partial u}}{{\partial x}}} \right|_{x = 0,x = l}} = 0, \hfill \\
  {\left. u \right|_{y = 0,y = h}} = 0,{\text{ }}{\left. {\frac{{\partial u}}{{\partial y}}} \right|_{y = 0,y = h}} = 0. \hfill \\ 
\end{gathered}
\end{equation}

In turn, the boundary conditions for the wedge with the free edges imply
\begin{equation}
\label{bc_free}
\begin{array}{l}
{\left. u \right|_{x = 0,x = l}} = 0,{\rm{ }}{\left. {\frac{{{\partial ^2}u}}{{\partial {x^2}}}} \right|_{x = 0,x = l}} = 0,\\
{\left. u \right|_{y = 0,y = h}} = 0,{\rm{ }}{\left. {\frac{{{\partial ^2}u}}{{\partial {y^2}}}} \right|_{y = 0,y = h}} = 0.
\end{array}
\end{equation}

It is important to emphasize that one has to distinguish between conditions $u=0$ for free edges and $u=0$ for the clamped edges. In the first case $u=0$ means that edges of a system under consideration rest on a fixed support but are not clamped to it. In the second case the condition $u=0$ expresses the fact that the edges of a wedge undergo no transversal displacement in the deformation, i.e., the edges are rigidly fixed.

We are searching for a solution in the form
\begin{equation}
\label{bc_free_sol1}
u(x,y,t) = U\left( {x,y} \right)\cos \left( {{\omega _t}t + \varphi } \right),
\end{equation}
that after the straightforward substitution Eq. (\ref{biharmonic}) transforms to stationary biharmonic equation
\begin{equation}
\label{bc_free_eq}
{\Delta ^2}U = {\kappa ^4}U,
\end{equation}
with ${\kappa ^4} = \frac{{12\left( {1 - {\nu ^2}} \right)\rho }}{{E{w^2}}}{\omega_{t}^2}$. Here $\omega_{t}$ is the eigenfrequency of transverse phonon modes and $\varphi$ is the arbitrary phase.

The case of vibration of the wedge with free edges (the boundary conditions are given by Eq. (\ref{bc_free})) allows the analytical
solution, which acquires the form:
\begin{equation}
\label{bc_free_sol2}
\begin{gathered}
  u\left( {x,y,t} \right) = A\left( {\sin \frac{{m\pi x}}{l}\sin \frac{{n\pi y}}{h} - \sin \frac{{n\pi x}}{l}\sin \frac{{m\pi y}}{h}} \right) \hfill \\
  \qquad {\text{                 }} \times \cos \left( {{\omega _t}t + \varphi } \right), \hfill \\ 
\end{gathered}
\end{equation}
where $A$ and $\varphi$ are the arbitrary amplitude and phase. The set of nonzero integer numbers $m$ and $n$ numerates the harmonics of the displacement field along the $x$- and $y$-axis respectively.
Corresponding eigenfrequencies for $m$ and $n$ with $m > n$ are given by the expression
\begin{equation}
\label{eigen_free}
{\omega _t} = {\pi ^2}\sqrt {\frac{E}{{12\rho \left( {1 - {\nu ^2}} \right)}}} \frac{w}{{{h^2}}}\left( {{m^2}{{\tan }^2}\theta  + {n^2}} \right).
\end{equation}
 
 Eq. (\ref{bc_free_eq}) for the wedge-shaped geometry under consideration is solved by subtracting of the two solutions of the same equation but for a rectangular membrane with the reversed indices $m$ and $n$. Since the displacement field on diagonals which is equidistant from the center must have the same expression (because of symmetry) this procedure gives a solution which vanishes along the diagonal as long as $m$ and $n$  are both even or odd.
 The lowest frequency corresponds to $m=3$ and $n=1$
\begin{equation}
\label{eigen_min_free}
\omega _t^{\left( {\min } \right)} = {\pi ^2}\sqrt {\frac{E}{{12\rho \left( {1 - {\nu ^2}} \right)}}} \frac{w}{{{h^2}}}\left( {9{{\tan }^2}\theta  + 1} \right).
\end{equation}

In the case of the wedge vibrations with clamped edges (boundary conditions are given by Eq. (\ref{bc_fixed})),  analytical solution of the biharmonic Eq. (\ref{biharmonic}) does not exist and numerical methods should be applied.  However, for the
elongated wedge, when $h \ll l$,  one can reduce the two-dimensional biharmonic equation to the quasi-one-dimensional one and find the solution of Eq. (\ref{bc_free_eq}) in the form \cite{Landau}
\begin{equation}
\label{bc_fixed_sol1}
\begin{gathered}
  u\left( {x,y,t} \right) = A\left[ {\left( {\sin \kappa l - \sinh \kappa l} \right)\left( {\cos \kappa x - \cosh \kappa x} \right)  } \right. \hfill \\
  \left. -{\left( {\cos \kappa l - \cosh \kappa l} \right)\left( {\sin \kappa x - \sinh \kappa x} \right)} \right]\cos \left( {\omega t + \varphi } \right), \hfill \\ 
\end{gathered}
\end{equation}
with the eigenfrequencies obeyed the transcendental equation
\begin{equation}
\label{eigen_fixed}
\cos \kappa (\omega_{t}) l\cosh \kappa (\omega_{t}) l = 1. 
\end{equation}
The minimal eigenfrequency can be found by means of the expansion of the left-side part of Eq. (\ref{eigen_fixed})
in series with respect to the  $\omega_{t}$ up to the third order of smallness. This gives
\begin{equation}
\label{eigen_fixed_min}
\omega _t^{\left( {\min } \right)} \approx \frac{{6w}}{{{l^2}}}\sqrt {\frac{E}{{\rho \left( {1 - {\nu ^2}} \right)}}}.
\end{equation}

\begin{figure}[ht]
\includegraphics[width=0.499\textwidth]{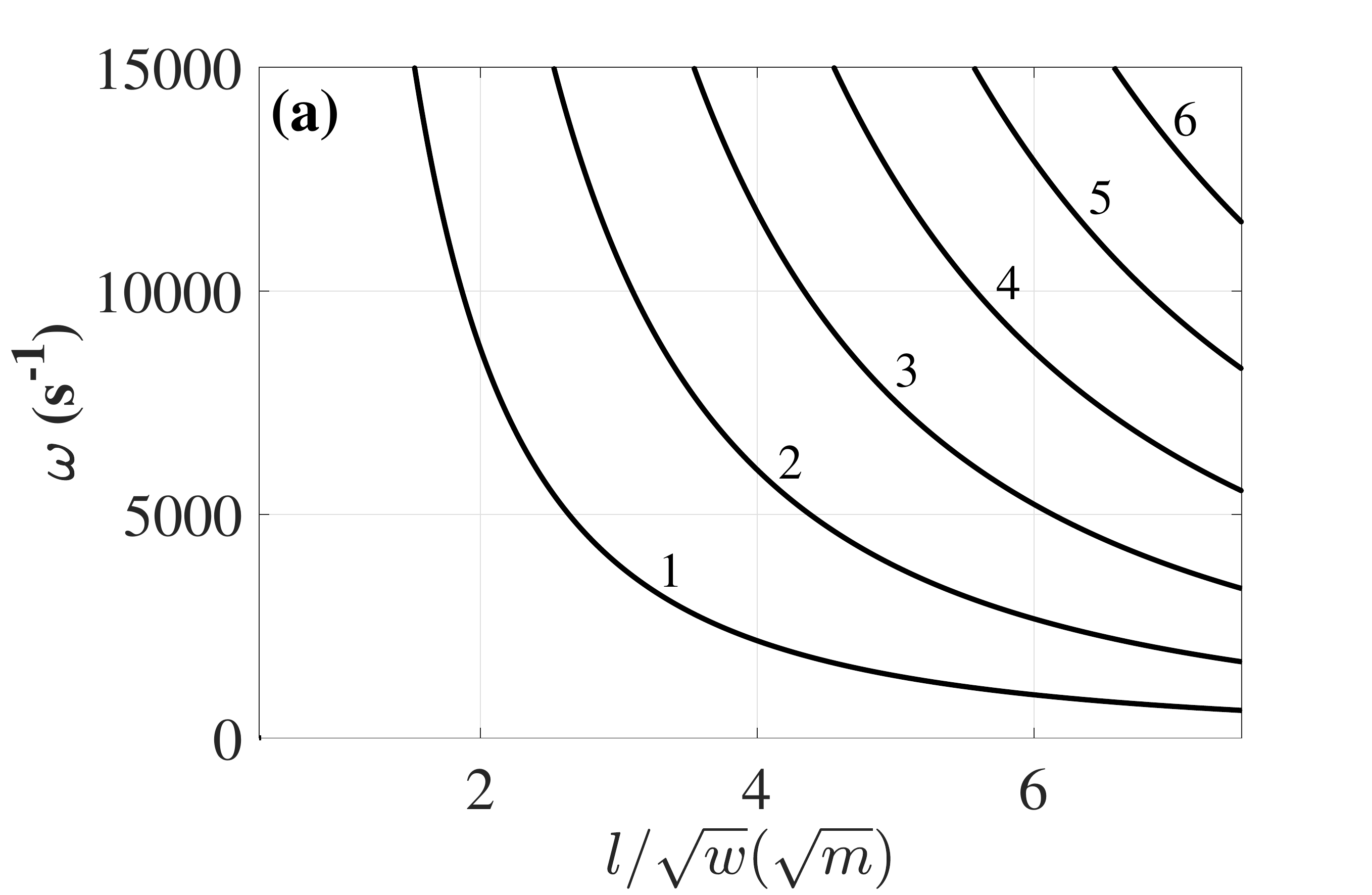}
\includegraphics[width=0.499\textwidth]{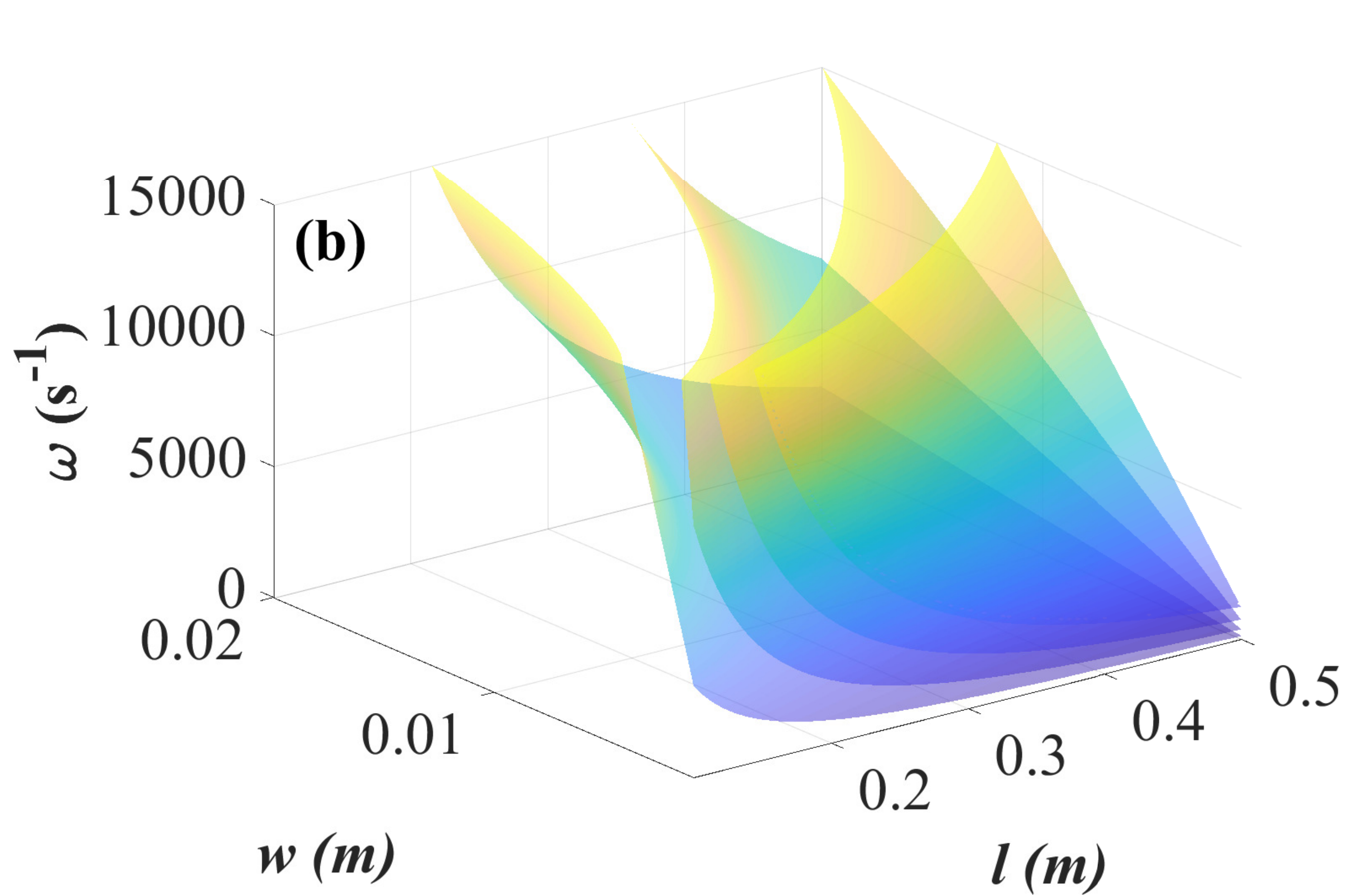}
\caption{Eigenfreqiencies of transverse phonon modes in a quartz wedge with the clamped edges when $h \ll l$ in a dependence of the length and the width. (a) Curves corresponding to different mode numbers are shown and numerated. (b) The three-dimensional plot of eigenfreqiencies dependence as a function of the length and the width of a wedge.}
\label{Eigen_trans}
\end{figure}

In Fig. 2 one can see the results of a numerical solution of Eq. (\ref{eigen_fixed}) for the eigenfrequencies as a function of combination its length and the square root of the width ($l/\sqrt w$) with the numerical values for parameters  $\rho = \SI{2650}{kg/m^3}$, $E=\SI{76.5} {GPa}$  and $\nu=0.07$ appropriate to the quartz (Fig. \ref{Eigen_trans}a). For the sake of visibility we plot a 3D dependence vs. geometrical parameters to show explicitly its dependence on the length and the width of a quartz wedge (Fig. \ref{Eigen_trans}b).

\subsection{Eigenfrequencies of longitudinal phonon modes}

If a wedge is no more thin and its width larger than phonon wavelength (see the section above) the longitudinal phonon modes are localized in a "bulky" system. In this case for the eigenfrequency problem the wave equation can be used
\begin{equation}
\label{harmonic}
\rho \frac{{{\partial ^2}u}}{{\partial {t^2}}} = \left( {\lambda  + 2\mu } \right)\Delta u,
\end{equation}
where $\lambda$ and $\mu$ are Lam\'e parameters.

Since we are interested in eigenfrequencies only, for both cases of clamped and free boundary conditions separated considerations are not required  \cite{footnote1}. To this end without loss of generality one can consider the case of fixed edges:
\begin{equation}
\label{bc_longitudinal}
{\left. u \right|_{x = 0,x = l}} = 0,{\rm{ }}{\left. u \right|_{y = 0,y = h}} = 0.
\end{equation}
The previously introduced constraint that the displacement field has only single component $u_x = u(x, y)$ should imply an additional boundary condition in the form of a certain stress at the inclined edge of the wedge to satisfy this assumption. However, as stated above, the subject of our consideration is the eigenfrequencies only, thus such a condition will not affect the result.
The solution of Eq. (\ref{harmonic}) with the boundary conditions Eq. (\ref{bc_longitudinal}) is similar to Eq. (\ref{bc_free_sol2})
\begin{equation}
\label{bc_fixed_sol2}
\begin{gathered}
  u\left( {x,y,t} \right) = A\left( {\sin \frac{{m\pi x}}{l}\sin \frac{{n\pi y}}{h} - \sin \frac{{n\pi x}}{l}\sin \frac{{m\pi y}}{h}} \right)  \hfill \\
 \qquad  \times \cos \left( {\omega t + \varphi } \right), \hfill \\ 
\end{gathered}
\end{equation}
where $\omega_{l}$ is the eigenfrequency of longitudinal phonon modes, $A$ and $\varphi$ are the arbitrary amplitude and phase and the set of nonzero integer numbers $m > n$ numerates the harmonics of $u(x,y,t)$ along the $x$- and $y$-axis respectively, like for the transverse modes (see subsection A above). However, in this case, the eigenfrequencies are given by the expression
\begin{equation}
\label{eigen_fixed_new}
{\omega _l} = \frac{\pi }{h}\sqrt {\frac{{{E}}}{\rho }} \sqrt {{m^2}{{\tan }^2}\theta  + {n^2}},
\end{equation}
that differs from Eq.(\ref{eigen_free}) for the eigenfrequencies of transverse phonon modes. 
The minimal frequency in this case is
\begin{equation}
\label{eigen_min_fixed}
\omega _l^{\left( {\min } \right)} = \frac{\pi }{h} \sqrt {\frac{E}{\rho }} \sqrt {9{{\tan }^2}\theta  + {1^2}}.
\end{equation}

\section{Dispersion relations for a phononic tsunami}

\subsection{The governing equation}

The occurrence of phononic tsunami in a wedge-shaped sample can be described by means of a nonlinear partial differential equation \cite{Kivshar}
\begin{widetext}
\begin{equation}
\label{govern_equation}
\begin{gathered}
  \frac{1}{{{c^2}}}\frac{{{\partial ^2}u}}{{\partial {t^2}}} = \frac{{{\partial ^2}u}}{{\partial {x^2}}} + \frac{{{\partial ^2}u}}{{\partial {y^2}}} + {\gamma _1}\left( {\frac{{{\partial ^4}u}}{{\partial {x^4}}} + \frac{{{\partial ^4}u}}{{\partial {y^4}}}} \right) + {\gamma _2}\frac{{{\partial ^4}u}}{{\partial {x^2}\partial {y^2}}}  \\ 
   + \alpha \left( {{{\left( {\frac{{\partial u}}{{\partial x}}} \right)}^2}\frac{{{\partial ^2}u}}{{\partial {x^2}}} + {{\left( {\frac{{\partial u}}{{\partial y}}} \right)}^2}\frac{{{\partial ^2}u}}{{\partial {y^2}}}} \right) + \beta \left( {{{\left( {\frac{{\partial u}}{{\partial y}}} \right)}^2}\frac{{{\partial ^2}u}}{{\partial {x^2}}} + 4 \frac{{\partial u}}{{\partial x}}\frac{{\partial u}}{{\partial y}}\frac{{{\partial ^2}u}}{{\partial x\partial y}} + {{\left( {\frac{{\partial u}}{{\partial x}}} \right)}^2}\frac{{{\partial ^2}u}}{{\partial {y^2}}}} \right). \\ 
\end{gathered}
\end{equation}
\end{widetext}
It was firstly introduced for the description of solitons in a nonlinear elastic medium and was derived in the frameworks of the microscopic scalar model considering the interacting particles with quartic polynomial potential. Within this approximation dimensionless coefficients $\alpha$ and $\beta$ can be interpreted as the re-scaled parameters that take into account the effect of anharmonicity, while dimensional coefficients $\gamma_1$ and $\gamma_2$ ($\dim {\gamma _i} = {{\rm{L}}^2}$) are responsible for the harmonic contribution to the potential. The coefficients $\gamma_1$ and $\gamma_2$ are chosen to be positive in order to stabilize the relative displacements between atoms of the medium from equilibrium. In turn $\alpha$ and $\beta$ can be positive or negative, resulting in additive or competitive contributions to the polynomial potential, correspondingly. The parameter $c$ can be considered as the longitudinal velocity of sound in a wedge-shaped sample. Also, within the derivation procedure $c$ plays role of the normalization factor in the definition of parameters $\alpha$, $\beta$, $\gamma_1$ and $\gamma_2$. The displacement field $\mathbf{u}$ has a single component that is directed along the $z$-axis and it does not depend on $z$, whereby the width of a wedge is not small in comparison with its length and height (see Figure 1).

It should be noted that Eq. (\ref{govern_equation}) does not include the terms that are responsible for the long-range interaction. The model under consideration and the corresponding Hamiltonian, which is the starting point for the derivation of Eq. (\ref{govern_equation}), stipulates a short-range elasticity only. This means that the emergence of the nonlinear phenomenon like a phononic tsunami starts within a single-connected component, created by neighboring particles of the elastic media. If one were use the Hamiltonian with long-range  kernel then one could trigger such a kind of instability in several connected components, spatially separated in the space that in the end can lead to even more stable nonlinear phenomenon manifestation. Therefore, results that will be discussed below remain the same on the qualitative level.

The presence of the higher derivatives of the displacement in Eq. (\ref{govern_equation}) may look rather surprising and the optional complication of the model. The emergence of a phonon tsunami implies the occurrence of sufficiently high strain levels in the wedge-shaped sample. In this case the continuum mechanics described by Eqs. (\ref{biharmonic}) and (\ref{harmonic}) fails and the discrete structure of a media should be taken into account. However, the discreteness of a crystal structure has the consequence that the relation between stress and strain in it can acquire a nonlocal character. Such a nonlocal relation between stress and strain leads to the spatial dispersion of phonon modes in the media. This results in the need to take into account higher order derivatives of the displacement field function. Using the continuum approximation when the crystallographic fractional coordinates can be considered as continuous variables, where the typical wavelength is much larger than the distance between the coupled objects and where no longer need to refer to the displacement of each interacting object, one can expand the displacement around the given atom in the lattice under consideration in Taylor series up to the fourth order and obtain  Eq. (\ref{govern_equation}) (see details of the derivation in Ref. \onlinecite{Kivshar}).

Moreover, the geometric nonlinearity of the system induced by the wedge-shaped sample dictates the strong necessity of the nonlinear potential for our model and consequently stresses out the importance of the higher order derivatives in Eq. (\ref{govern_equation}) for the description of a phononic tsunami in such a geometry.

It is noteworthy that Eq. (\ref{govern_equation}) can be derived also in the framework of nonlinear elasticity theory \cite{Maradudin2}. Using the long wave limit of the equation of motion of a simple crystal with nearest and next nearest neighbor central and non-central force interactions between atoms, equations of motion for the stress tensor can be expanded in series up to fourth derivative terms for the displacement vector components.

When $ {\gamma _1}={\gamma _2} =0 $ and $ \alpha  = \beta  = 0$ (i. e. the lack of anharmonicity),  we arrive to the well-known wave equation Eq. (\ref{harmonic}).

It is worth note also that recently the simplified version of Eq.  (\ref{govern_equation}) with $\alpha  = 0$, $\beta  = 0$ and ${\gamma _2} = 2{\gamma _1}$ was used for the prediction of new class of phononic metamaterials, in which the phonon band dispersion can be changed from an acoustic to an optical type by modulating a uniform stress \cite{Karki}. In this work it was demonstrated theoretically how to stop and switch signals in a tunable metamaterial by changing the dispersion relation of an entire band.

\subsection{Dispersion relations}

Based on the hydrodynamic-like behavior of phonons (see Introduction) one can extend to our model the concept of the Airy wave linear theory. It is well known that the latter is the cornerstone for the derivation of the dispersion relation for an ocean tsunami at the intermediate depth. To avoid confusions, we would like to recall that this hydrodynamic similarity is based on the formal analogy between an ocean tsunami and its phononic counterpart. The constitutive equation is still Eq. (\ref{govern_equation}), obtained within the nonlinear elasticity theory. Based on this approach we calculate dispersion relations of a phononic tsunami far away from the narrow edge of a wedge-shaped sample. Under such a condition, the nonlinear effects induced by the geometry of a system can be neglected and anharmonic effects are vanishing $\alpha=\beta=0$.  This implies the linear structure of the wave packet along the x-axis and allows to seek the solution of Eq. (\ref{govern_equation}) in the form:
\begin{equation}
\label{govern_equation_sol}
u\left( {x,y,t} \right) = {u_0}\left( y \right){e^{ikx - i\omega t}}.
\end{equation}
Substitution of Eq. (\ref{govern_equation_sol}) into Eq. (\ref{govern_equation}) yields an ordinary differential equation
\begin{equation}
\label{govern_equation_1D}
{\gamma _1}\frac{{{d^4}{u_0}}}{{d{y^4}}} + \left( {1 - {k^2}{\gamma _2}} \right)\frac{{{d^2}{u_0}}}{{d{y^2}}} + \left( {{\gamma _1}{k^4} + \frac{{{\omega ^2}}}{{{c^2}}} - {k^2}} \right){u_0} = 0,
\end{equation}
which solution is
\begin{equation}
\label{govern_equation_sol_1D}
{u_0}\left( y \right) = {C_1}{e^{ - {q_1}y}} + {C_2}{e^{{q_1}y}} + {C_3}{e^{ - {q_2}y}} + {C_4}{e^{{q_2}y}},
\end{equation}
where $C_i$ are constants and
\begin{widetext}
\begin{equation}
\label{q1}
{q_1} = \frac{1}{2}\frac{{\sqrt {2{\gamma _1}\left( {{\gamma _2}{k^2} - 1 - \sqrt { - 4\gamma _1^2{k^4} + \gamma _2^2{k^4} - 4{\gamma _1}\frac{{{\omega ^2}}}{{{c^2}}} + 4{\gamma _1}{k^2} - 2{\gamma _2}{k^2} + 1} } \right)} }}{{{\gamma _1}}},
\end{equation}
\begin{equation}
\label{q2}
{q_2} = \frac{1}{2}\frac{{\sqrt {2{\gamma _1}\left( {{\gamma _2}{k^2} - 1 + \sqrt { - 4\gamma _1^2{k^4} + \gamma _2^2{k^4} - 4{\gamma _1}\frac{{{\omega ^2}}}{{{c^2}}} + 4{\gamma _1}{k^2} - 2{\gamma _2}{k^2} + 1} } \right)} }}{{{\gamma _1}}}.
\end{equation}
\end{widetext}

\subsubsection{Clamped edges}
The boundary conditions for the case of clamped edges given by Eq. (\ref{bc_fixed}) are transformed to a more amenable form:
\begin{equation}
\label{bc_fixed_y}
\begin{gathered}
  {u_0}\left| {_{y = 0}} \right. = {\left. {{u_0}} \right|_{y = h}} = 0, \hfill \\
  {\left. {\frac{{d{u_0}}}{{dy}}} \right|_{y = 0}} = {\left. {\frac{{d{u_0}}}{{dy}}} \right|_{y = h}} = 0. \hfill \\ 
\end{gathered}
\end{equation}
Taking into account boundary conditions Eq. (\ref{bc_fixed_y})  one can rewrite the solution of Eq. (\ref{govern_equation_sol_1D})  in the form
\begin{equation}
\label{sol_1D_fixed}
\begin{gathered}
  {u_0}\left( y \right) = C\left[ {\sinh \left( {{q_1}\left( {y - h} \right)} \right) + \sinh {q_1}h\cosh {q_2}y} \right.   \hfill \\
  \qquad {\text{             }}\left. +{\sinh {q_1}y\cosh {q_2}h} \right]{q_2} + C\left[ {\sinh \left( {{q_2}\left( {y - h} \right)} \right)} \right.   \hfill \\
 \qquad  {\text{            }}\left. +{\sinh {q_2}y\cosh {q_1}h + \sinh {q_2}h\cosh {q_1}y} \right]{q_1}, \hfill \\ 
\end{gathered} 
\end{equation}
where $C$ is an arbitrary constant. 

The dispersion relation $\omega=\omega(k)$ for a phononic tsunami is determined by the solvability condition for the constants $C_i$ in Eq. (\ref{govern_equation_sol_1D}) and is given by the implicit expression
\begin{equation}
\label{dispersion_fixed_new}
\begin{gathered}
  2{q_1}{q_2}\left( {1 - \frac{1}{{\cosh \left( {{q_1}h} \right)\cosh \left( {{q_2}h} \right)}}} \right)   \hfill \\
\qquad = \left( {q_1^2 + q_2^2} \right)\tanh \left( {{q_1}h} \right)\tanh \left( {{q_2}h} \right), \hfill \\ 
\end{gathered}
\end{equation}
where $\omega$ and $k$ are included in Eqs. (\ref{q1}) and (\ref{q2}).
It is interesting to note that, due to the presence of the hyperbolic tangent function, the dispersion relation given by Eq. (\ref{dispersion_fixed_new}) formally reminds of the analogous ''simplified'' characteristic of a tsunami wave in the ocean ${\omega ^2} = gk\tanh \left( {kh} \right)$ (see Introduction). 

\begin{figure}[ht]
\centering
\includegraphics[width=0.495\textwidth]{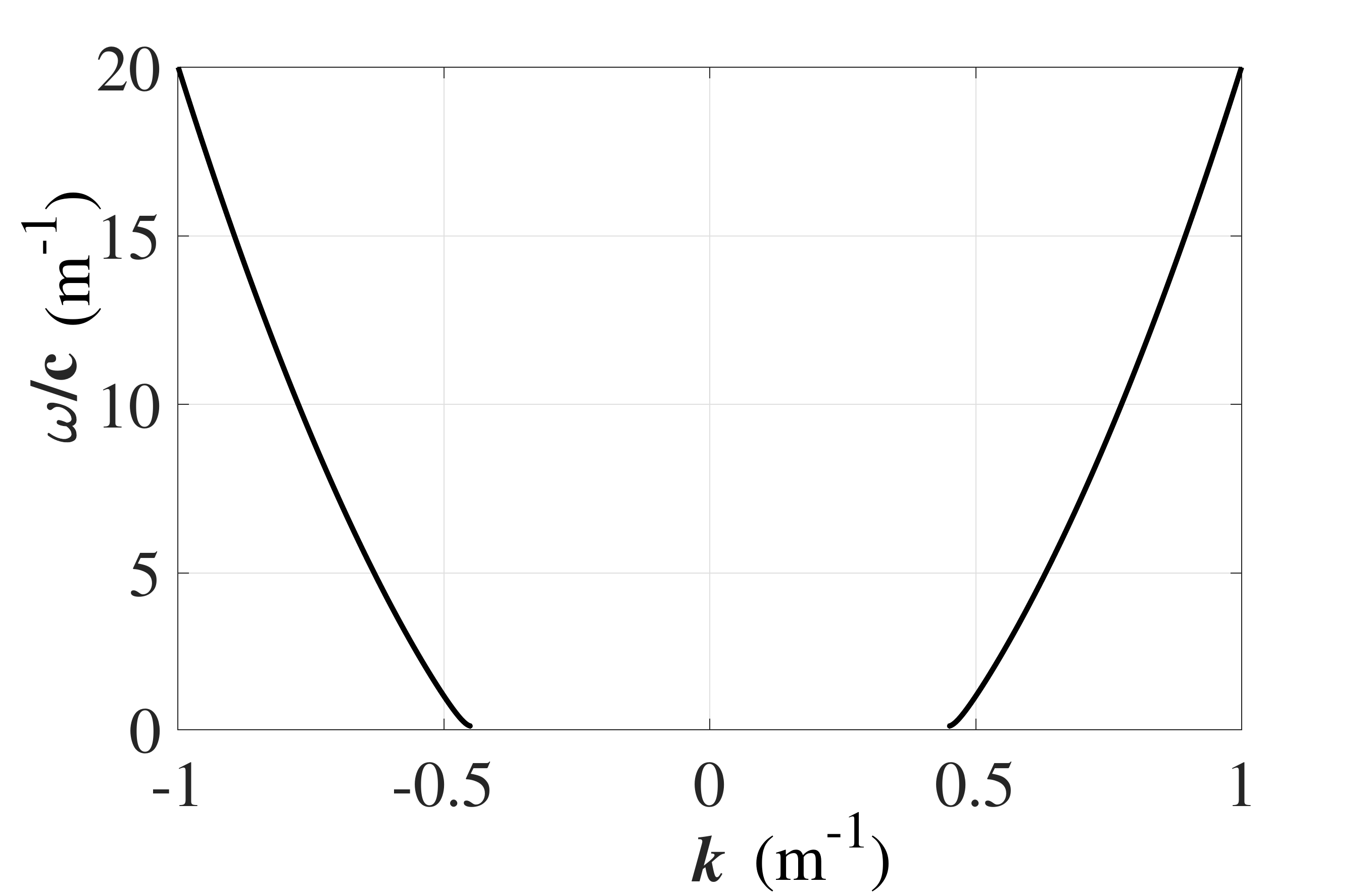}
\caption{The dispersion relation for a phononic tsunami in a wedge with clamped edges for ${\gamma _1} = \SI{0.01} {m^2}$, ${\gamma _2} = \SI{5} {m^2}$ and $h=\SI{0.05} {m}$.}
\label{dispersion_clamped}
\end{figure}

The numerical solution of Eq. (\ref{dispersion_fixed_new}) is shown in Figure \ref{dispersion_clamped}. The remarkable feature of the dispersion relation for the given set of parameters is the appearance of  a gap in momentum space. Such a kind of the dispersion relation with $\textit{k}$-gap is not a new one and was predicted for different liquids and supercritical fluids \cite{Yang}, plasma \cite{Kalman, Murillo} and for certain holographic models of quantum field theory \cite{Baggioli, Trachenko}. Direct experimental evidence for the $\textit{k}$-gap has been obtained for the phonon spectra in a monolayer dusty plasma \cite{Nosenko}. Therefore, we can consider gapped momentum states in corresponding experiments as a fingerprint of the phononic tsunami. 

\subsubsection{Free edges}
In the case of free edges, where the boundary conditions
\begin{equation}
\label{bc_free_y}
\begin{gathered}
  {u_0}\left| {_{y = 0}} \right. = {\left. {{u_0}} \right|_{y = h}} = 0, \hfill \\
  {\left. {\frac{{d^2{u_0}}}{{dy^2}}} \right|_{y = 0}} = {\left. {\frac{{d^2{u_0}}}{{dy^2}}} \right|_{y = h}} = 0. \hfill \\ 
\end{gathered}
\end{equation}
are imposed, the $y$-component of the displacement field is described by the function
\begin{equation}
\label{sol_1D_free}
\begin{gathered}
  {u_0}\left( y \right) = C\left[ {\sinh \left( {{q_1}\left( {y - h} \right)} \right) + \sinh {q_1}h\cosh {q_2}y} \right.   \hfill \\
 \qquad {\text{             }}\left. +{\sinh {q_1}y\cosh {q_2}h} \right]{q_2^2} + C\left[ {\sinh \left( {{q_2}\left( {y - h} \right)} \right)} \right.   \hfill \\
  \qquad {\text{            }}\left. +{\sinh {q_2}y\cosh {q_1}h + \sinh {q_2}h\cosh {q_1}y} \right]{q_1^2} \hfill \\ 
\end{gathered}
\end{equation}
where $C$ is an arbitrary constant. 
\begin{figure}[ht]
\centering
\includegraphics[width=0.495\textwidth]{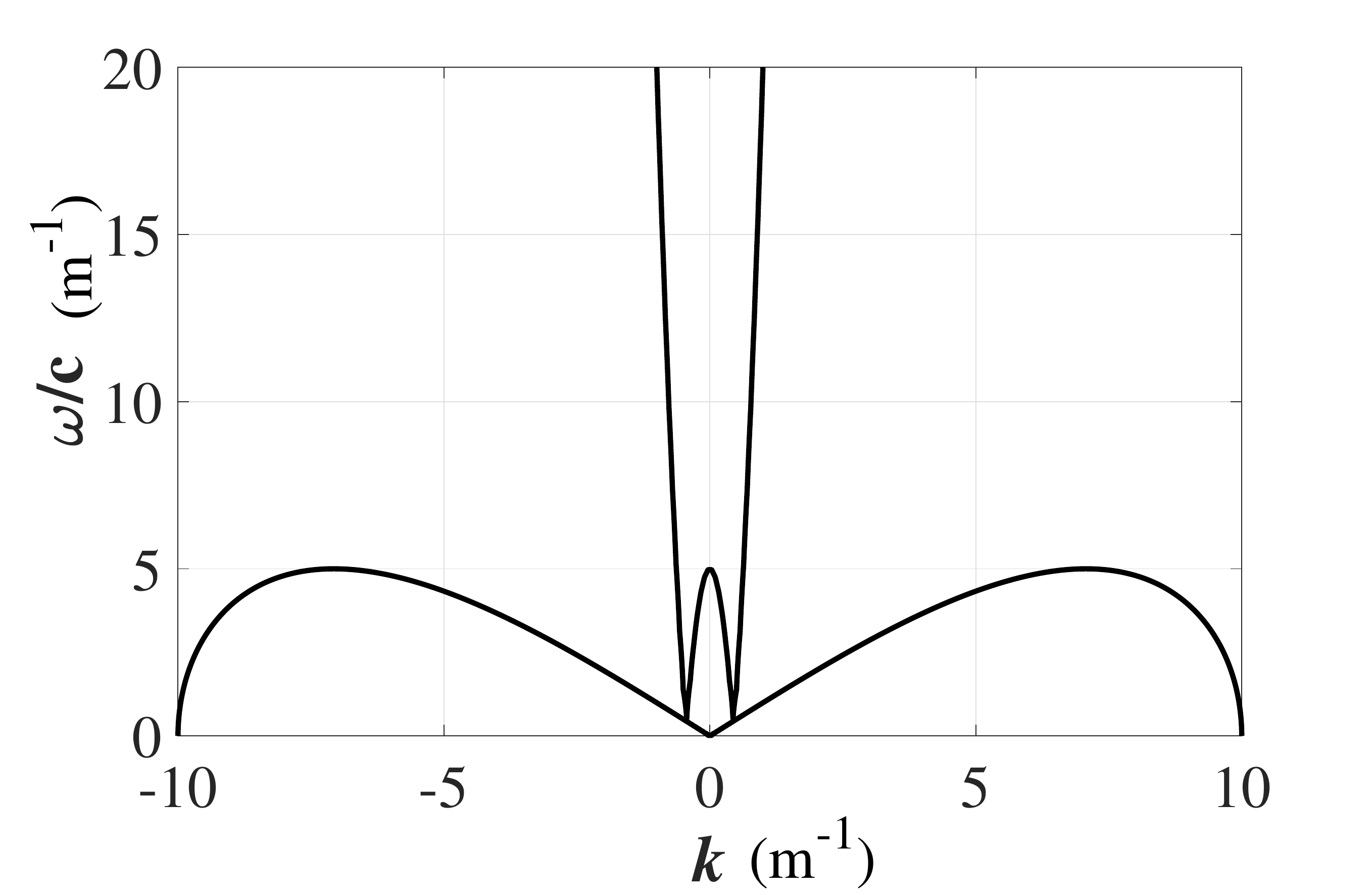}
\caption{The dispersion relation for a phononic tsunami in a wedge with free edges for ${\gamma _1} = \SI{0.01} {m^2}$, ${\gamma _2} = \SI{5} {m^2}$ and $h=\SI{0.05} {m}$.}
\label{dispersion_free}
\end{figure}
The dispersion relation has the form of a multi-valued function that assumes two distinct values for the given value of $k$
\begin{equation}
\label{dispersion_free_new}
\left\{ \begin{array}{l}
\frac{{{\omega ^2}}}{{{c^2}}} = \frac{{\left( {\gamma _2^2 - 4\gamma _1^2} \right){k^4} - 2\left( {{\gamma _2} - 2{\gamma _1}} \right){k^2} + 1}}{{4{\gamma _1}}},\\
\frac{{{\omega ^2}}}{{{c^2}}} = \left( {1 - \gamma _1^2{k^2}} \right){k^2}.
\end{array} \right.
\end{equation}

It is worth noting that at qualitative level Eq. (\ref{dispersion_free_new}) is nothing else than the first two terms in the series expansion of the dispersion relation of the ocean tsunami ${\omega ^2} = gk\tanh \left( {kh} \right)$ (see Introduction).
The dispersion relation given by Eq. (\ref{dispersion_free_new}) is shown in the Figure \ref{dispersion_free}.

Another striking feature of our analytical consideration is that the $y$-component of the displacement amplitudes represented by Eqs.  (\ref{sol_1D_fixed}) and (\ref{sol_1D_free}) are similar to the expression for the deviation of the water surface from the undisturbed state, obtained within Airy wave theory for a ocean tsunami (see e.g. Ref. \onlinecite{Torsvik}).

\section{Numerical results for a stationary case}

\begin{figure*}
\includegraphics[width=1\columnwidth]{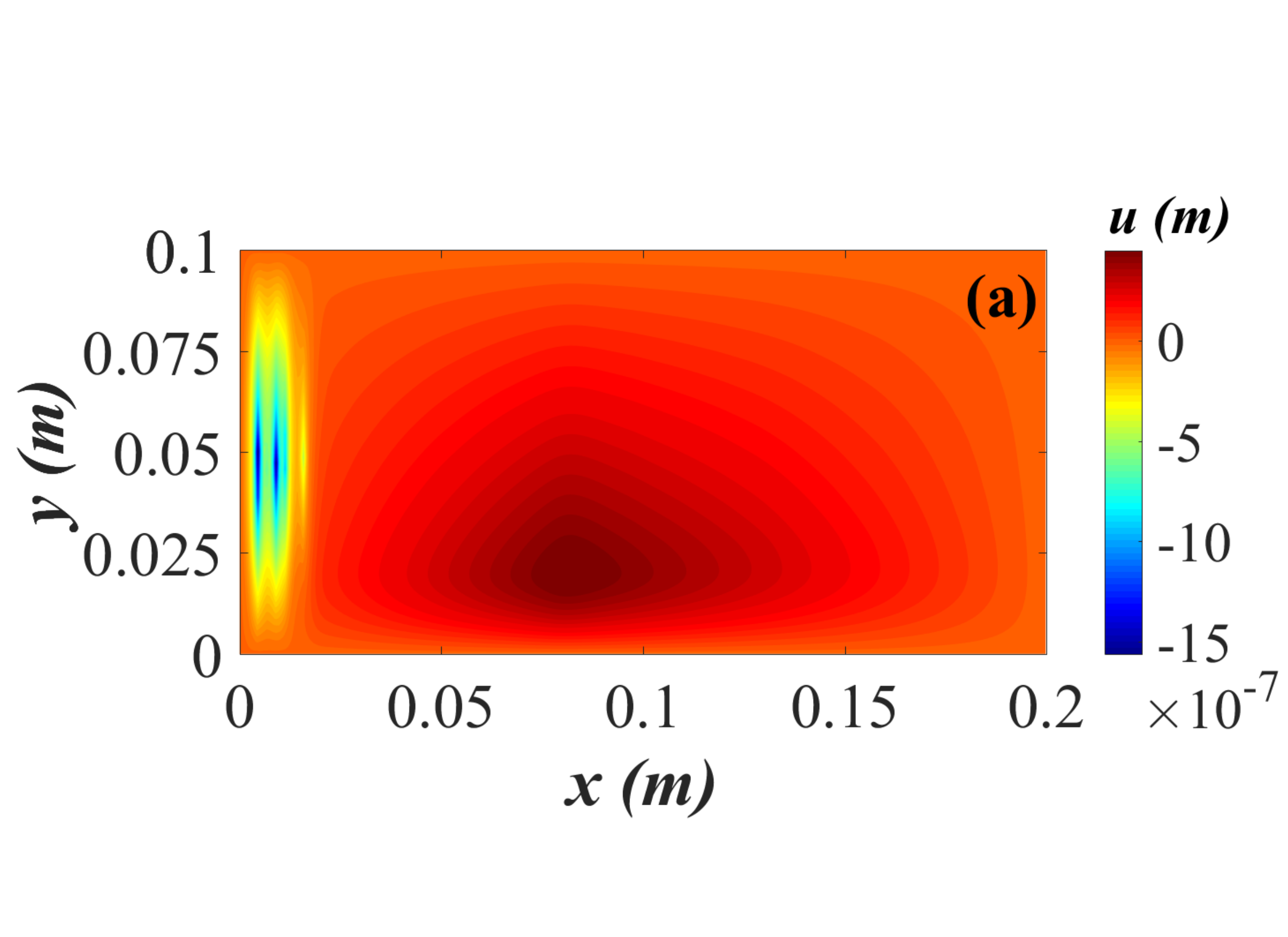}
\includegraphics[width=1\columnwidth]{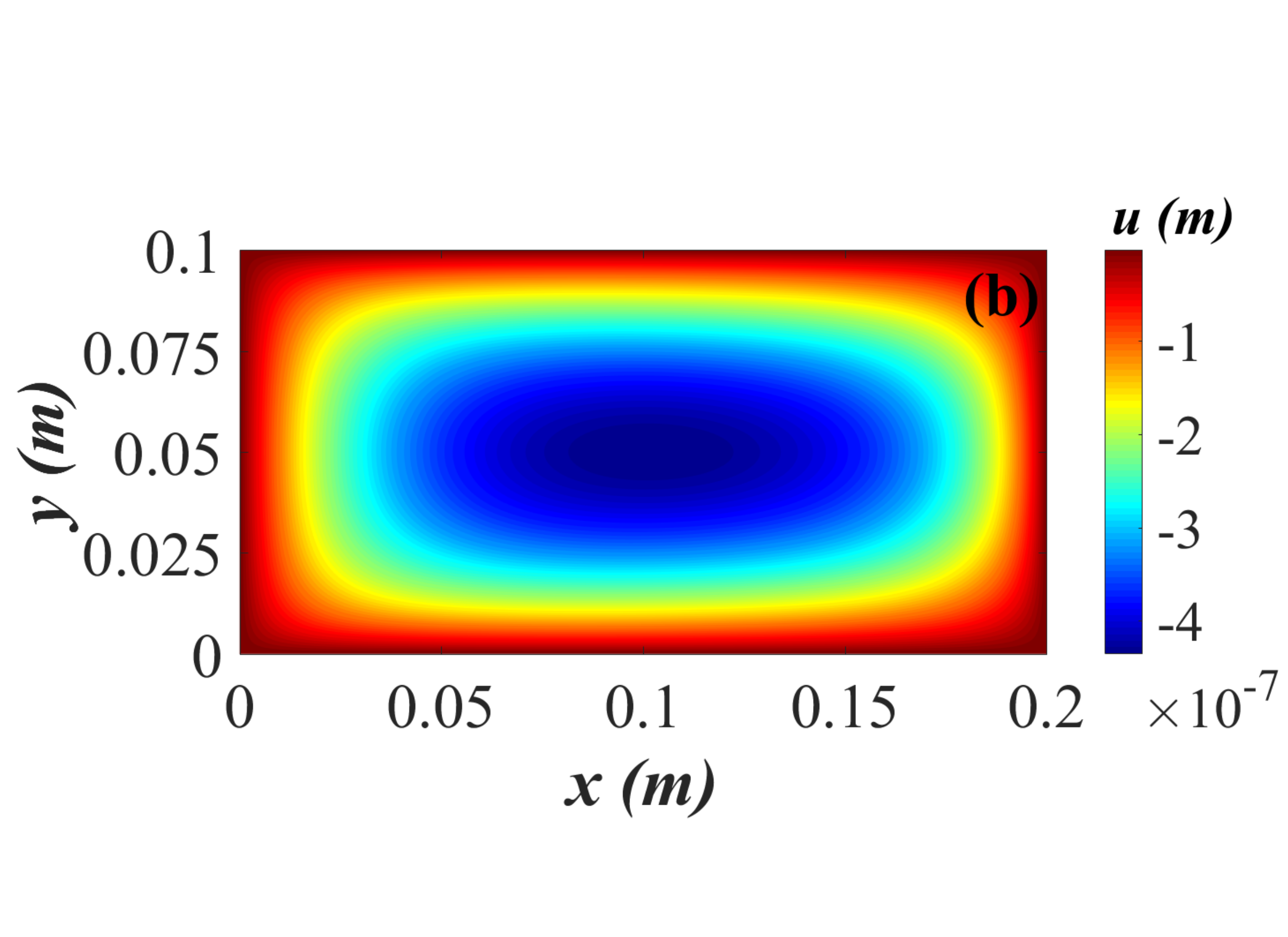}
\includegraphics[width=1\columnwidth]{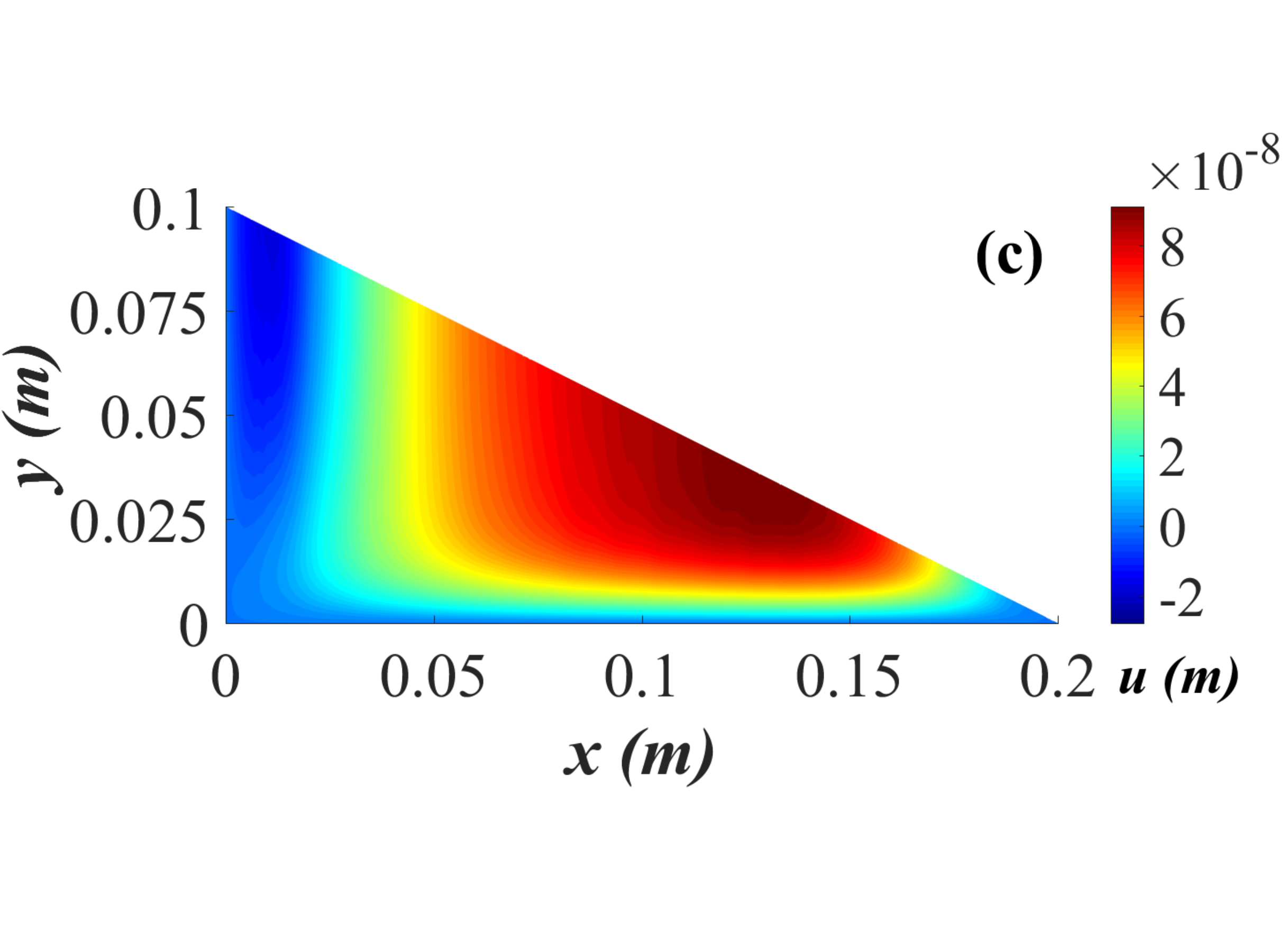}
\includegraphics[width=1\columnwidth]{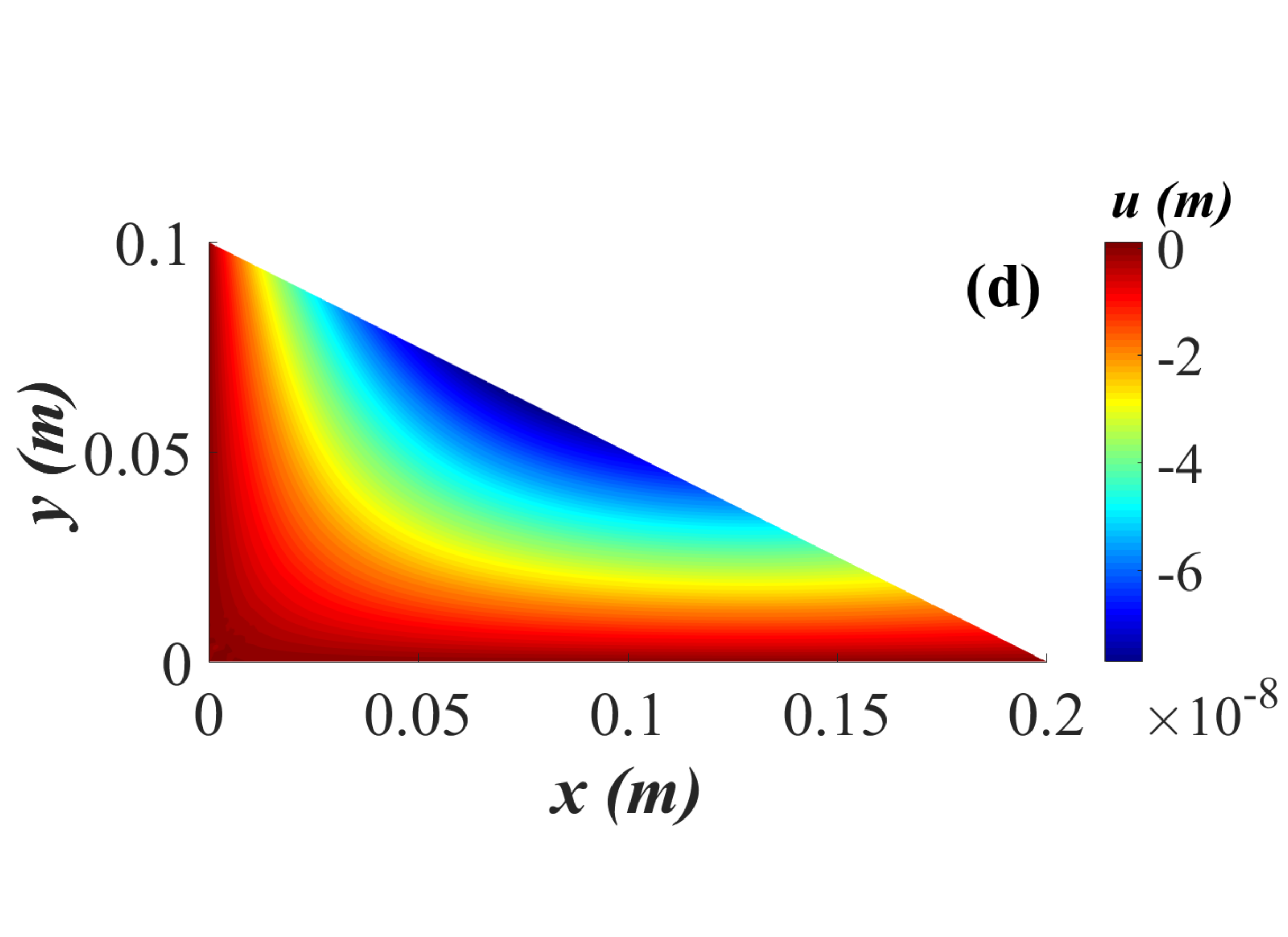}
\caption{Displacement amplitudes in a rectangular cuboid (a, b) and a wedge (c,d) with the height $h=\SI{0.1} {m}$ and the length $l=\SI{0.2} {m}$ for the set of dimensionless parameters $\gamma_1=\SI{0.01} {m^2}$, $\gamma_2=\SI{5}{m^2}$, $\alpha=0$, $\beta=0$ (a, c) and  $\gamma_1=\SI{0.01} {m^2}$, $\gamma_2=\SI{5} {m^2}$, $\alpha=1$, $\beta=1$ (b, d).}
\label{stat_numerical}
\end{figure*}

To justify our predictions we performed numerical calculations of the stationary  Eq. (\ref{govern_equation}) for boundary conditions corresponding to free edges to show the possibility of realizing of a phononic tsunami in a wedge-shaped sample:
\begin{widetext}
\begin{equation}
\label{govern_equation_stat}
\begin{gathered}
  \frac{{{\partial ^2}u}}{{\partial {x^2}}} + \frac{{{\partial ^2}u}}{{\partial {y^2}}} + {\gamma _1}\left( {\frac{{{\partial ^4}u}}{{\partial {x^4}}} + \frac{{{\partial ^4}u}}{{\partial {y^4}}}} \right) + {\gamma _2}\frac{{{\partial ^4}u}}{{\partial {x^2}\partial {y^2}}}  \\ 
   + \alpha \left( {{{\left( {\frac{{\partial u}}{{\partial x}}} \right)}^2}\frac{{{\partial ^2}u}}{{\partial {x^2}}} + {{\left( {\frac{{\partial u}}{{\partial y}}} \right)}^2}\frac{{{\partial ^2}u}}{{\partial {y^2}}}} \right) + \beta \left( {{{\left( {\frac{{\partial u}}{{\partial y}}} \right)}^2}\frac{{{\partial ^2}u}}{{\partial {x^2}}} + 4 \frac{{\partial u}}{{\partial x}}\frac{{\partial u}}{{\partial y}}\frac{{{\partial ^2}u}}{{\partial x\partial y}} + {{\left( {\frac{{\partial u}}{{\partial x}}} \right)}^2}\frac{{{\partial ^2}u}}{{\partial {y^2}}}} \right) = 0. \\ 
\end{gathered}
\end{equation}
\end{widetext}

Due to highly nonlinearity of Eq. (\ref{govern_equation_stat}) we cannot apply the same procedure with the subtraction of two solutions with indices reversed as it was performed for Eqs. (\ref{biharmonic}) and (\ref{harmonic}) to satisfy the boundary conditions for the inclined wedge edge. Generally speaking, the boundary conditions should be written in the following way
\begin{equation}
\label{general_bc_numerical}
\frac{{{\partial ^2}u}}{{\partial {{\mathbf{n}}^2}}} = 0,
\end{equation}
where $d/d\mathbf{n}$ denote differentiation along the outward normal to the slope of a wedge-shaped sample \cite{Landau}. However, according to our numerical simulations the imposing of Eq. (\ref{general_bc_numerical}) lead to significant increase of the calculation time. To avoid such a problem and speed up numerical solution the inclined edge is treated as the stress free (see details of the numerical simulation in Appendix A). This assumption does not change essentially the results on qualitative and quantitative level and preserves our further conclusions (see below) regardless of the boundary conditions selection for the inclined edge of a wedge.

First of all we show the importance of the system geometry and the shoaling effect for the observation of a phononic tsunami. To this end we solved numerically  Eq. (\ref{govern_equation_stat}) for a rectangular cuboid with the same dimensions as in the case of a wedge. Asymmetric structure of a displacement field in Figure  \ref{stat_numerical}a is connected to the significant difference between coefficients $\gamma_1=\SI{0.01} {m^2}$ and $\gamma_2=\SI{5} {m^2}$ and the simultaneous lack of the ''stabilizing'' parameters $\alpha=0$ and $\beta=0$ and, as a consequence, an amplification of nonlinear effects. Introduction of  $\alpha=1$ and $\beta=1$ leads to the smoothing on nonlinear effects (Fig.  \ref{stat_numerical}b) and the transformation to the symmetric shape of the displacement field. 

In the case of a wedge-shaped sample one can see from Figures  \ref{stat_numerical}(c) and  \ref{stat_numerical}(d) that there are solutions with the significant enhancement of the displacement in a wedge, which can be attributed to the emergence of a tsunami phononic wave.  
Moreover, Figure  \ref{stat_numerical} (c) and (d) show the amplitude of $u$ approximately \SI{10} {nm}, which agrees with the experimental results for the peak value of the displacement arising from of phonon solitons occurrence in a 1D phononic crystal waveguide\cite{Kurosu}. Justification of the observed data in the above-mentioned experiment was done within the dynamic Euler-Bernoulli equation (Euler-Bernoulli beam theory) with the subsequent reduction to the nonlinear Schrodinger equation. It is interesting to note that under certain conditions Eq. (\ref{govern_equation}) also can be transformed to the nonlinear Schrodinger equation.

\section{Dynamical behavior of a phononic tsunami induced by a short Gaussian pulse excitation: analytical approach}

Before we restricted our study to the consideration of  the stationary Eq. (\ref{govern_equation_stat}). We have shown the possibility of generation of the topological excitation in the form of a phononic tsunami and have provided detectable signatures of this effect. Until now we excluded from investigation the dynamics of this phenomenon in a wedge-shaped sample. 
However, from the theoretical point of view it can be done by solution of the time-dependent Eq. (\ref{govern_equation}) with introduced of a short Gaussian pulse $f(t) = \sqrt{\frac{\ln{2}}{\pi}} \, \frac{f_0}{t_{1\!/2}} \exp{\Bigg[- \frac{t^2 \ln{2}}{t_{1\!/2}^2}\Bigg]}$ to the right side of Eq. (\ref{govern_equation}). The parameter $t_{1\!/2}$ is the pulse half width at half maximum and $f_0$ is its strength, applied to the vertical edge of the wedge. 

When $t_{1\!/2}$ is close to zero one can induce a phononic tsunami. The vanishing value of $t_{1\!/2}$ mimics a sudden displacement of the ocean needed for the generation of tsunami wave. Very recently, the similar strategy was proposed for the inducing another interesting topological effect in phononic systems. By means of appropriately tuned AC acoustic sources located on the boundary of the body the dissipative-free motion of lattice defects can be observed \cite{Gorbushin}. 

To illustrate the phononic tsunami emergence together with its dynamical evolution and to avoid extremely extensive numerical simulations of the time-dependent Eq. (\ref{govern_equation}), one can proceed to the analytical consideration. With this in mind, we adopt several reasonable assumptions. Firstly, due to a wedge-shaped geometry for the imitation of the shoaling effect of the tsunami wave we expect the dramatic growth of the displacement amplitude near the narrow edge of the horizontal side. Secondly, within formulated theoretical approach one can assume the initial existence of the nonlinear wave excitation in a system. Using the similarity of the ocean tsunami behavior near the coast these two assumptions allow to determine the boundary conditions for Eq. (\ref{govern_equation}) for the right edge of the wedge and put ${\left. u \right|_{x = l}}$ tends to be large and ${\left. {\frac{{{d^2}u}}{{d{x^2}}}} \right|_{x = l}} \to 0$. From the physical point of view such conditions correspond to the significant enhancement of the amplitude with the decrease of the curvature of the displacement field. The latter is equivalent to the ``vertical sea wall'' of the ocean tsunami near the coastline. Finally, the last assumption is that the rest of boundaries are treated as the stress free.

These simplifications allow to naively decompose the displacement field $u(x,y)$ on $x$ and $y$-axis independently, using the ansatz
\begin{equation}
\label{anzatz}
u\left( {x,y,t} \right) = A{u_0}\left( x \right) + B\left( {y - {V_g}t} \right) + F\left( t \right).
\end{equation}
Here $A$ and $B$ are the amplitudes of the displacement field along the $x$- and $y$-axis respectively and ${V_g} \equiv \frac{{\partial \omega }}{{\partial k}}$ is the group velocity.  The function $F\left( t \right)$
\begin{equation}
\label{F}
\begin{gathered}
  F\left( t \right) = \frac{1}{2}{f_0}{t_{1/2}}{c^2}\left( {\frac{t}{{{t_{1/2}}}}{\text{erf}}\left( {\sqrt {\ln 2} \frac{t}{{{t_{1/2}}}}} \right)} \right. \hfill \\
 \qquad \left. { + \frac{1}{{\sqrt {\pi \ln 2} }}\exp \left( { - \ln 2\frac{{{t^2}}}{{t_{1/2}^2}}} \right)} \right), \hfill \\ 
\end{gathered}
\end{equation}
represents the result of double integration of $f(t)$ over $t$, where ${\text{erf}}\left( z \right)$  is the error function.
The specifics of this anzats is that a tsunami wave arises ``from nothing'', following the terminology of Ref.  \onlinecite{Lighthill}. The existence of such an class of solutions is not new and was introduced, for instance, for the construction of the Boussinesq equation \cite{Kaptsov} for the description of shallow-water waves.

By means of the ansatz Eq. (\ref{anzatz}) we proceed to the ordinary nonlinear differential equation of the 4th order instead of Eq. (\ref{govern_equation}):
\begin{equation}
\label{u0}
{\gamma _1}\frac{{{d^4}{u_0}}}{{d{x^4}}} + \alpha {A^2}\frac{{{d^2}{u_0}}}{{d{x^2}}}{\left( {\frac{{d{u_0}}}{{dx}}} \right)^2} + \left( {1 + \beta {B^2}} \right)\frac{{{d^2}{u_0}}}{{d{x^2}}} = 0.
\end{equation}
The ansatz  Eq. (\ref{anzatz}) eliminates in Eq. (\ref{govern_equation}) the term with the coefficient $\gamma_2$. Roughly speaking, it means that the interaction between atoms of the elastic media along $x$-axis is significantly stronger than along $y$ and $z$ directions. Simultaneously to preserve the total contributions of other coefficients we suppose that second-, third-, and fourth-order elastic constants are nonzero and, therefore, cannot be ignored. 

Eq. (\ref{u0}) can be solved in terms of the incomplete elliptic integral of the third kind  $\Pi (z,\nu, m)$ and the elliptic Jacobi function ${\text{sn}}\left( {z,m} \right)$. The list of solutions with the details of derivation are given in Appendix B. The choice of an appropriate solution is governed solely by the behavior of $u_0(x)$.  The plot of this function must contain pronounced spikes, identified with the unique feature of tsunami known as the dramatic increase of the amplitude of the wave near the coastline. As it is shown in the Appendix B the selection procedure is based on the behavior of the composite function $\Pi \left( {{\text{sn}}\left( {x,m} \right),\nu ,m} \right)$. For the certain interval of $x$ the  $u_0(x)$ acquires a significant enhancement only in the case when the characteristic of the elliptic integral of the third kind $\nu$ is larger than zero. It is this criterion that is used to select solution to describe of a phononic tsunami dynamics. 

Based on this criterion and choosing the corresponding expression for $u_0(x)$ one obtains
\begin{equation}
\label{u0_alpha<0_1}
u_0(x)=\frac{1}{\lambda }\left( {{r_1} - {r_2}} \right)\Pi \left( {{\text{sn}}\left( {\lambda x,q'} \right),\frac{{{r_1} - {r_4}}}{{{r_2} - {r_4}}},q'} \right) + {r_2}x.
\end{equation}
Here ${r_1} \geqslant {r_2} \geqslant 0 \geqslant {r_3} \geqslant {r_4}$ are the real roots of the depressed quartic equation (see Appendix B and Eq. (\ref{polynomial})), $\lambda={\frac{1}{2}\sqrt {\frac{{\left| \alpha  \right|}}{{6{\gamma _1}}}} \sqrt {\left( {{r_1} - {r_3}} \right)\left( {{r_2} - {r_4}} \right)} }$ and the modulus of the Jacobi elliptic sine $q' = \sqrt {1 - {q^2}}$ with $q = \sqrt {\frac{{\left( {{r_1} - {r_2}} \right)\left( {{r_3} - {r_4}} \right)}}{{\left( {{r_1} - {r_3}} \right)\left( {{r_2} - {r_4}} \right)}}}$. Figure \ref{u0} shows the plot of the function $u_0(x)$ with the characteristic pronounced growth at the narrow edge of a wedge. 
\begin{figure}
\includegraphics[width=1\columnwidth]{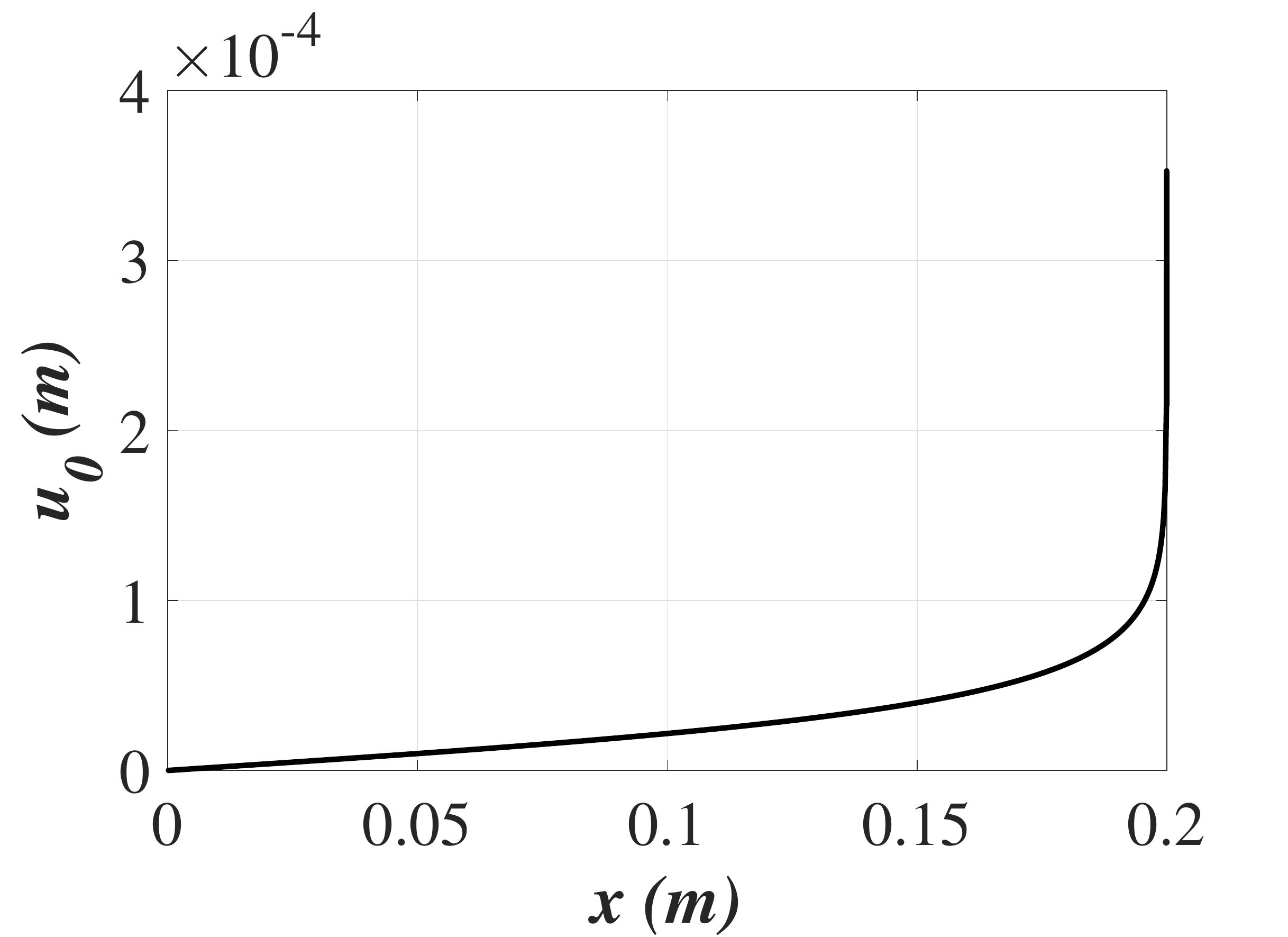}
\caption{The plot of the function $u_0(x) $ with the dramatic enhancement at the narrow edge of a wedge $x=l$, which imitates the ``vertical wall'' of a ocean tsunami wave. The set of parameters are $\gamma_1=\SI{0.18} {m^2}$, $\alpha  =  - 4.5 \cdot {10^8}$ and $\beta={10^8}$, also ${V_g} \approx c = \SI{5800} {m/s}$.}
\label{u0}
\end{figure}

Therefore, the final expression for the function $u$ takes the form
\begin{equation}
\label{u_final}
\begin{gathered}
  u(x,y,t) = \frac{A}{\lambda }\left( {{r_1} - {r_2}} \right)\Pi \left( {{\text{sn}}\left( {\lambda x,q'} \right),\frac{{{r_1} - {r_4}}}{{{r_2} - {r_4}}},q'} \right) \\ 
   + {r_2}x + B\left( {y - V_g t} \right) + F\left( t \right). \\ 
\end{gathered}
\end{equation}

\begin{figure}
\includegraphics[width=1\columnwidth]{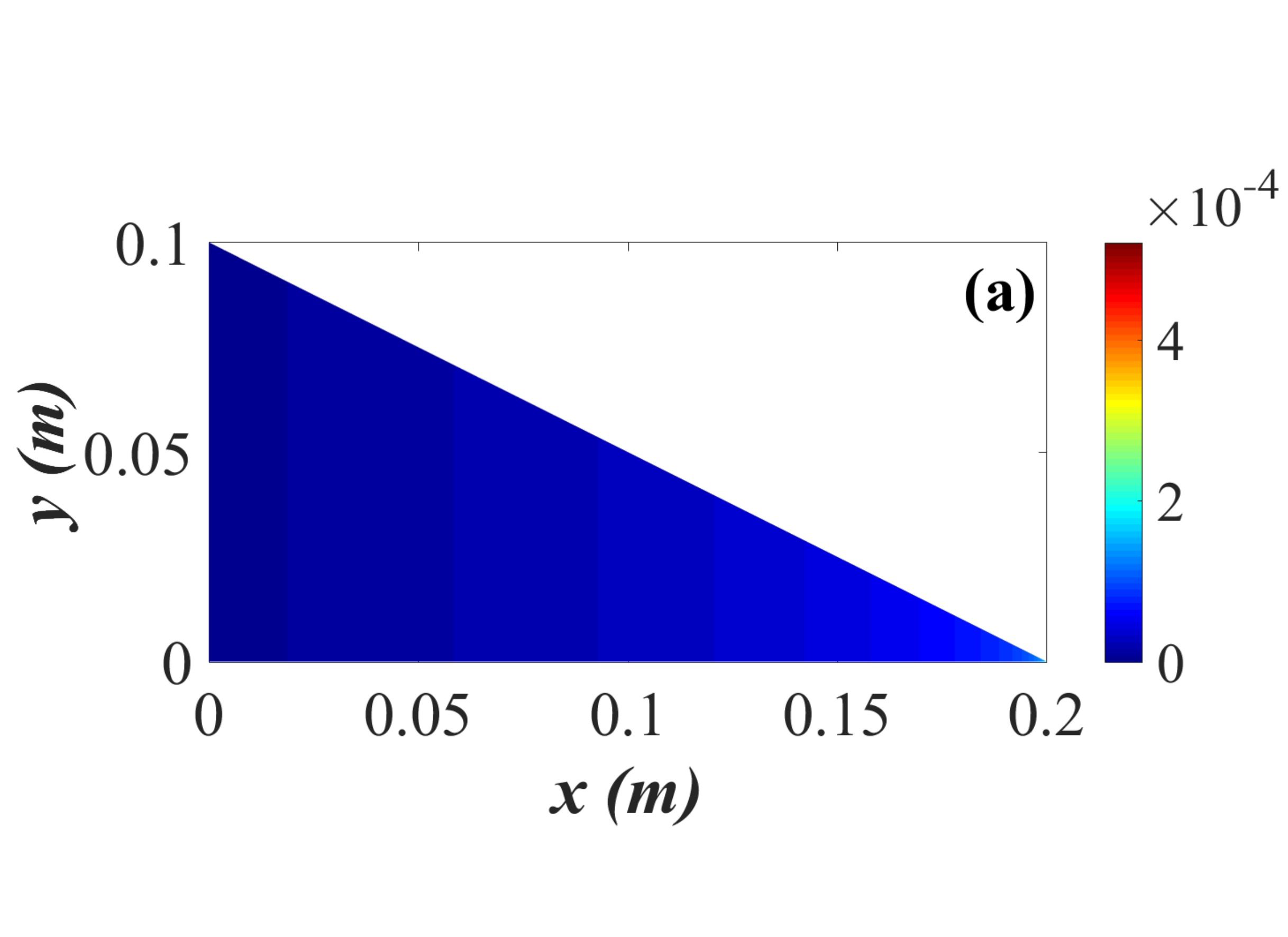}
\vspace{0.00mm}
\includegraphics[width=1\columnwidth]{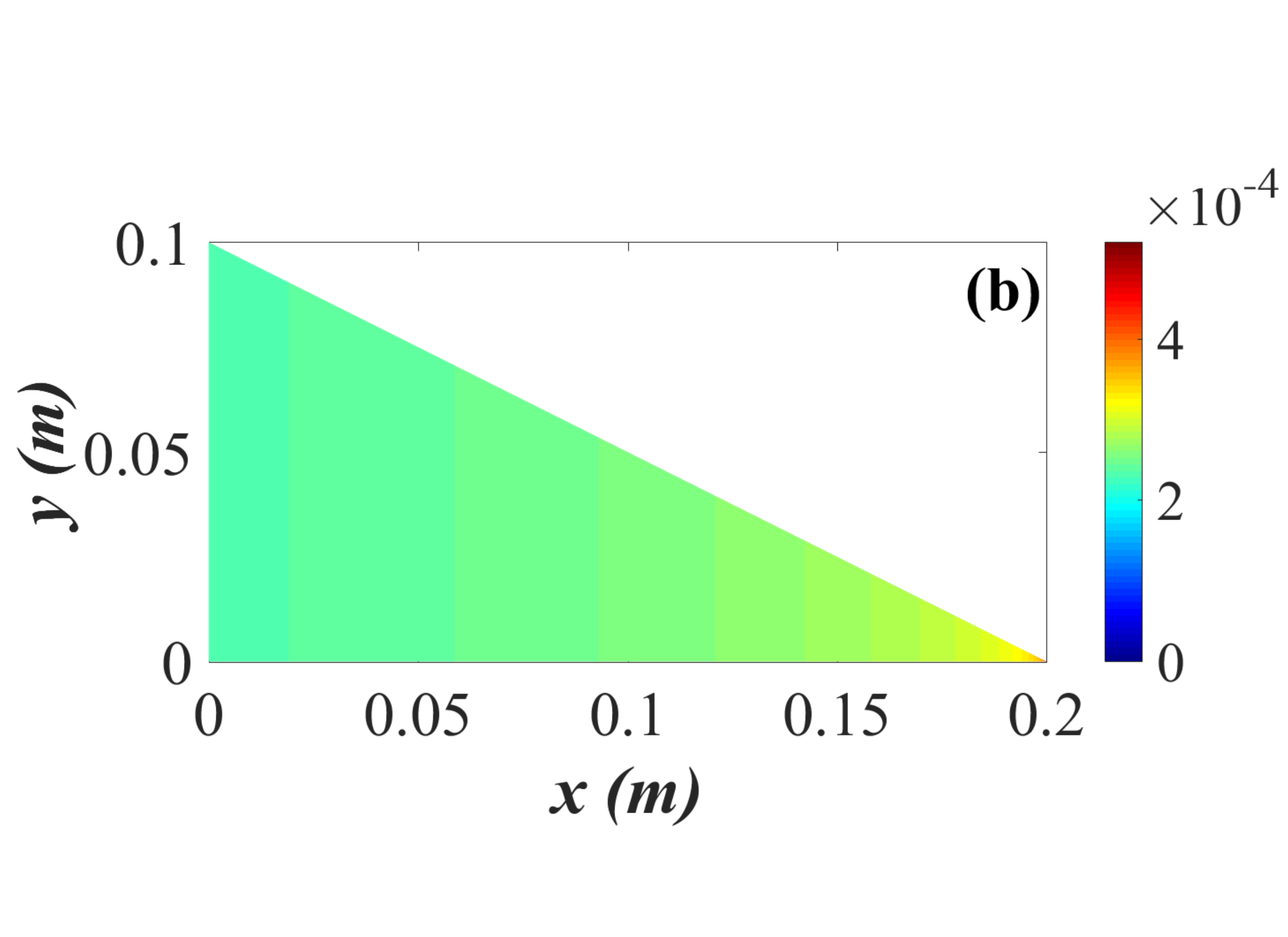}
\caption{The dynamics of the displacement field in a wedge-shaped sample induced by  Gaussian pulse with $t_{1\!/2} = \SI{1} {\mu s}$ and the amplitude $f_0 =  4 \cdot {10^{-7}} \SI {} {m^{-1}s} $ at the initial moment of time (a) and at $t = \frac{l}{c} \approx \SI{34.48} {\mu s}$ (b) for the set of parameters of the elastic media  $\gamma_1=\SI{0.18} {m^2}$, $\alpha  =  - 4.5 \cdot {10^8}$ and $\beta={10^8}$.}
\label{u_dynamics}
\end{figure}

Eq. (\ref{u_final}) shows the dynamical behavior of the displacement field in a wedge that can be attributed to a phononic tsunami occurrence, namely the shoaling effect with the significant increase in the amplitude at the narrow edge (see Fig. \ref{u_dynamics}). Comparing displacement fields at the initial moment of time $t=0$ shown in Figure \ref{u_dynamics} (a) and after applying the input Gaussian pulse at $t=l/c$ in Figure \ref{u_dynamics} (b) one can see the peak value of $u \approx  6 \cdot {10^{ - 4}}{\text{ m}}$ in the narrow edge of a wedge-shaped sample. This value is much larger than obtained within numerical simulations in Figure \ref{stat_numerical} (c) and (d), where the maximum of the displacement reaches $\SI{10} {nm}$.

The advantage of the analytical solution given by Eq. (\ref{u_final}) is that it can be used for the description and the fitting of data in future experiments on the detection of a phononic tsunami phenomenon in a more effective way without extensive time-dependent numerical simulations of Eq. (\ref{govern_equation}).

\section{Conclusions}

In this article the way of realization of a phononic tsunami in a crystal lattice as a novel nonlinear phenomenon is proposed. The  experimental conditions for the observation of such an effect are also delineated.  We outline the particular potential of wedge-shaped samples use to generation of phononic tsunami waves. The dispersion relations that have been calculated within a theory of solitons in a nonlinear elastic medium can be considered as a hallmark of a phononic tsunami. The occurrence of such a kind of nonlinear phononic excitations in wedge-shaped geometry has been demonstrated by numerical simulations and by the analytical calculations. Such predictions can be verified by means of measurements of the displacement field induced by Gaussian short pulse in relevant materials including phononic crystals and acoustic metamaterials. Another experimental possibility is the determination of the dispersion relations predicted as a signature of the phononic tsunami in corresponding systems. 

\section{Acknowledgements}

This work was supported by the CarESS project.

\appendix
\section{Details of the finite-element analysis}
The numerical simulation of Eq. (\ref{govern_equation_stat}) was performed in COMSOL Multiphysics. For the solution of the fourth-order partial differential equation the general form PDE module was applied. 

For further implementation of the finite-element method we rewrite Eq. (\ref{govern_equation_stat}) in the form
\begin{widetext}
\begin{equation}
\label{govern_equation_comsol}
\begin{gathered}
  \frac{\partial }{{\partial x}}\left[ {\frac{{\partial u}}{{\partial x}} + {\gamma _1}\frac{\partial }{{\partial x}}\left( {\frac{{{\partial ^2}u}}{{\partial {x^2}}}} \right) + {\gamma _2}\frac{\partial }{{\partial x}}\left( {\frac{{{\partial ^2}u}}{{\partial {y^2}}}} \right) + \frac{1}{3}\alpha {{\left( {\frac{{\partial u}}{{\partial x}}} \right)}^3} + \beta \frac{{\partial u}}{{\partial x}}{{\left( {\frac{{\partial u}}{{\partial y}}} \right)}^2}} \right] +  \hfill \\
  \frac{\partial }{{\partial y}}\left[ {\frac{{\partial u}}{{\partial y}} + {\gamma _1}\frac{\partial }{{\partial y}}\left( {\frac{{{\partial ^2}u}}{{\partial {y^2}}}} \right) + \frac{1}{3}\alpha {{\left( {\frac{{\partial u}}{{\partial y}}} \right)}^3} + \beta \frac{{\partial u}}{{\partial y}}{{\left( {\frac{{\partial u}}{{\partial x}}} \right)}^2}} \right] = 0. \hfill \\ 
\end{gathered}
\end{equation}
\end{widetext}

Introducing new functions $P = \frac{{{\partial ^2}u}}{{\partial {x^2}}}$  and $Q = \frac{{{\partial ^2}u}}{{\partial {y^2}}}$  we reduce Eq. (\ref{govern_equation_comsol}) to the system of differential equations of the second order 
\begin{equation}
\label{govern_equation_comsol_new}
\left\{ \begin{gathered}
  \frac{\partial }{{\partial x}}\left[ {\frac{{\partial u}}{{\partial x}} + {\gamma _1}\frac{{\partial P}}{{\partial x}} + {\gamma _2}\frac{{\partial Q}}{{\partial x}} + \frac{1}{3}\alpha {{\left( {\frac{{\partial u}}{{\partial x}}} \right)}^3} + \beta \frac{{\partial u}}{{\partial x}}{{\left( {\frac{{\partial u}}{{\partial y}}} \right)}^2}} \right] \\ 
   + \frac{\partial }{{\partial y}}\left[ {\frac{{\partial u}}{{\partial y}} + {\gamma _1}\frac{{\partial Q}}{{\partial y}} + \frac{1}{3}\alpha {{\left( {\frac{{\partial u}}{{\partial y}}} \right)}^3} + \beta \frac{{\partial u}}{{\partial y}}{{\left( {\frac{{\partial u}}{{\partial x}}} \right)}^2}} \right] = 0, \\ 
  P = \frac{{{\partial ^2}u}}{{\partial {x^2}}}, \\ 
  Q = \frac{{{\partial ^2}u}}{{\partial {y^2}}}, \\ 
\end{gathered}  \right.
\end{equation}
which is adapted now for the numerical solution in COMSOL Multiphysics. 

The Dirichlet boundary conditions $u = 0$, $P=0$ and $Q=0$ were imposed for a wedge except the inclined edge that is treated as the stress free. 

A predefined mesh calibrated for fluid dynamics was used for the simulation procedure. Element size parameters for this mesh are as follows: maximum element size $= 0.0028$, minimum element size $=4 \cdot {10^{ - 5}}$, maximum element growth rate $= 1.1$, curvature factor $= 0.25$, and resolution of narrow regions $=1$. The convergence criterion is set to ${10^{ - 6}}$.

\section{Analytical solutions}

We redefine  $\tilde \alpha  \equiv \alpha {A^2}$ and $\tilde \beta  \equiv \beta {B^2}$ and rewrite Eq. (\ref{u0}) in the main text of the paper, omitting tilde 
\begin{equation}
\label{u01}
\frac{d}{{dx}}\left( {{\gamma _1}\frac{{{d^3}{u_0}}}{{d{x^3}}} + \left( {1 + \beta } \right)\frac{{d{u_0}}}{{dx}} + \frac{1}{3}\alpha {{\left( {\frac{{d{u_0}}}{{dx}}} \right)}^3}} \right) = 0,
\end{equation}
or
\begin{equation}
\label{u02}
{\gamma _1}\frac{{{d^3}{u_0}}}{{d{x^3}}} + \left( {1 + \beta } \right)\frac{{d{u_0}}}{{dx}} + \frac{1}{3}\alpha {\left( {\frac{{d{u_0}}}{{dx}}} \right)^3} = {C_1},
\end{equation}
where $C_1$ is an arbitrary constant.
To solve Eq. (\ref{u02}) the substitution  $\frac{{d{u_0}\left( x \right)}}{{dx}} = \psi$ is applied that gives
\begin{equation}
\label{u03}
{\gamma _1}\frac{{{d^2}\psi }}{{d{x^2}}} + \left( {1 + \beta } \right)\psi  + \frac{1}{3}\alpha {\psi ^3} = {C_1}.
\end{equation}

Since Eq. (\ref{u03}) does not depend explicitly on the coordinate  $x$ one can use a new change of variables $\frac{{d\psi }}{{dx}} = \eta$  such as $\frac{{{d^2}\psi }}{{d{x^2}}} = \eta \frac{{d\eta }}{{d\psi }}$
\begin{equation}
\label{u04}
{\gamma _1}\eta \frac{{d\eta }}{{d\psi }} + \left( {1 + \beta } \right)\psi  + \frac{1}{3}\alpha {\psi ^3} = {C_1}.
\end{equation}

The first integral is
\begin{equation}
\label{u05}
\frac{1}{2}{\gamma _1}{\eta ^2} + \frac{1}{2}\left( {1 + \beta } \right){\psi ^2} + \frac{1}{{12}}\alpha {\psi ^4} = {C_1}\psi  + {C_2},
\end{equation}
where $C_2$ is another arbitrary constant.

For $\alpha>0$ Eq. (\ref{u05}) transforms to
\begin{equation}
\label{u06}
\frac{{6{\gamma _1}}}{{\left| \alpha  \right|}}{\left( {\frac{{d\psi }}{{dx}}} \right)^2} + {\psi ^4} + \frac{{6\left( {1 + \beta } \right)}}{{\left| \alpha  \right|}}{\psi ^2} - 12{C_1}\psi  - 12{C_2} = 0,
\end{equation}
and for $\alpha<0$ we have
\begin{equation}
\label{u07}
 - \frac{{6{\gamma _1}}}{{\left| \alpha  \right|}}{\left( {\frac{{d\psi }}{{dx}}} \right)^2} + {\psi ^4} - \frac{{6\left( {1 + \beta } \right)}}{{\left| \alpha  \right|}}{\psi ^2} + 12{C_1}\psi  + 12{C_2} = 0.
\end{equation}

Let us assume that ${r_1}$, ${r_2}$, ${r_3}$ and ${r_4}$  are the real roots of the depressed quartic equation for $\alpha>0$ and $\alpha<0$ respectively
\begin{equation}
\label{polynomial}
\begin{gathered}
   {\psi ^4} + \frac{{6\left( {1 + \beta } \right)}}{{\left| \alpha  \right|}}{\psi ^2} - 12{C_1}\psi  - 12{C_2} = 0, \hfill \\
    {\psi ^4} - \frac{{6\left( {1 + \beta } \right)}}{{\left| \alpha  \right|}}{\psi ^2} + 12{C_1}\psi  + 12{C_2} = 0 \hfill \\ 
\end{gathered}
\end{equation}
that is satisfied the system of equations
\begin{equation}
\label{roots}
\begin{gathered}
  {r_1} + {r_2} + {r_3} + {r_4} = 0, \hfill \\
  {r_1}{r_2} + {r_1}{r_3} + {r_1}{r_4} + {r_2}{r_3} + {r_2}{r_4} + {r_3}{r_4} =  \pm \frac{{6\left( {1 + \beta } \right)}}{{\left| \alpha  \right|}}, \hfill \\
  {r_1}{r_2}{r_3} + {r_1}{r_2}{r_4} + {r_1}{r_3}{r_4} + {r_2}{r_3}{r_4} =   \mp  12{C_1}, \hfill \\
  {r_1}{r_2}{r_3}{r_4} =  \mp  12{C_2}. \hfill \\ 
\end{gathered}
\end{equation}

\subsection{$\alpha>0$}
We proceed to solve Eq. (\ref{u06}) putting the coefficient $\alpha$ larger than zero. Assuming all roots of Eq. (\ref{polynomial}) are real and have the order ${r_1} \geqslant {r_2} \geqslant 0 \geqslant {r_3} \geqslant {r_4}$ the solution of Eq. (\ref{u06}) is expressed explicitly via  the Jacobi elliptic sine function $\text{s}\text{n}(z)$
\begin{equation}
\label{psi_solution1}
\psi  = {r_4} + \frac{{\left( {{r_1} - {r_4}} \right)\left( {{r_2} - {r_4}} \right)}}{{\left( {{r_2} - {r_4}} \right) + \left( {{r_1} - {r_2}} \right){\text{s}}{{\text{n}}^2}\left( {\lambda x,q} \right)}},
\end{equation}
where $\lambda={\frac{1}{2}\sqrt {\frac{{\left| \alpha  \right|}}{{6{\gamma _1}}}} \sqrt {\left( {{r_1} - {r_3}} \right)\left( {{r_2} - {r_4}} \right)} }$ and the modulus of the Jacobi elliptic sine $q = \sqrt {\frac{{\left( {{r_1} - {r_2}} \right)\left( {{r_3} - {r_4}} \right)}}{{\left( {{r_1} - {r_3}} \right)\left( {{r_2} - {r_4}} \right)}}}$. The subsequent integration of Eq. (\ref{psi_solution1}) yields the final expression for function $u_0(x)$
\begin{equation}
\label{u0_solution1}
u_0(x)=\frac{1}{\lambda }\left( {{r_1} - {r_4}} \right)\Pi \left( {{\text{sn}}\left( {\lambda x,q} \right), - \frac{{{r_1} - {r_2}}}{{{r_2} - {r_4}}},q} \right) + {r_4}x,
\end{equation}
where $\Pi \left( {z,\nu, m} \right)$ the incomplete elliptic integral of the third kind.
   
For the case of the roots ordering ${r_1} \geqslant {r_2}  \geqslant {r_3} \geqslant {r_4} \geqslant 0$ Eq. (\ref{u06}) has another solution
\begin{equation}
\label{psi_solution2}
\psi  = {r_2} - \frac{{\left( {{r_2} - {r_3}} \right)\left( {{r_2} - {r_4}} \right)}}{{\left( {{r_2} - {r_4}} \right) - \left( {{r_3} - {r_4}} \right){\text{s}}{{\text{n}}^2}\left( {\lambda x,q} \right)}},
\end{equation}
and correspondingly 
\begin{equation}
\label{u0_solution2}
u_0(x)=-\frac{1}{\lambda }\left( {{r_2} - {r_3}} \right)\Pi \left( {{\text{sn}}\left( {\lambda x,q} \right), \frac{{{r_3} - {r_4}}}{{{r_2} - {r_4}}},q} \right) + {r_2}x.
\end{equation}

When Eq. (\ref{polynomial}) has the two real roots such as ${r_1} \geqslant {r_2}$ and two complex-conjugate roots $r_{3,4}=\rho_1 \pm i\rho_2$ one obtain
\begin{equation}
\label{psi_solution3}
\psi  = {r_1} - \frac{{\left( {{r_1} - {r_2}} \right)\left( {1 - {\text{cn}}\left( {\lambda x,q} \right)} \right)}}{{1 + \delta  - \left( {1 - \delta } \right){\text{cn}}\left( {\lambda x,q} \right)}},
\end{equation}
where we introduce another set of parameters $\delta = \frac{{\sqrt {{{\left( {{r_2} - {\rho _1}} \right)}^2} + \rho _2^2} }}{{\sqrt {{{\left( {{r_1} - {\rho _1}} \right)}^2} + \rho _2^2} }}$, the coefficient $\lambda  = \sqrt {\frac{{\left| \alpha  \right|}}{{6{\gamma _1}}}} {\left[ {\left( {{{\left( {{r_1} - {\rho _1}} \right)}^2} + \rho _2^2} \right)\left( {{{\left( {{r_2} - {\rho _1}} \right)}^2} + \rho _2^2} \right)} \right]^{\frac{1}{4}}}$ and the modulus of the Jacobi elliptic cosine $q = \sqrt {\frac{1}{2} - \frac{{\left( {{r_1} - {\rho _1}} \right)\left( {{r_2} - {\rho _1}} \right) + \rho _2^2}}{{\sqrt {\left( {{{\left( {{r_1} - {\rho _1}} \right)}^2} + \rho _2^2} \right)\left( {{{\left( {{r_2} - {\rho _1}} \right)}^2} + \rho _2^2} \right)} }}}$. The straightforward integration of Eq. (\ref{psi_solution3}) gives
\begin{widetext}
\begin{equation}
\label{u0_solution3}
\begin{gathered}
  {u_0}\left( x \right) = \left( {{r_1} + \frac{{{r_1} - {r_2}}}{{1 - \delta }}} \right)x - \frac{{\left( {{r_1} - {r_2}} \right)}}{{2\lambda }}\frac{{1 + \delta }}{{1 - \delta }}\left[ {\Pi \left( {{\text{sn}}\left( {\lambda x,q} \right), - \frac{{{{\left( {1 - \delta } \right)}^2}}}{{4\delta }},q} \right)} \right. \hfill \\
  \left. {\left. { + \frac{{1 - \delta }}{{1 + \delta }}\sqrt {\frac{{1 - {{\left( {\frac{{1 - \delta }}{{1 + \delta }}} \right)}^2}}}{{{q^2} + \left( {1 - {q^2}} \right){{\left( {\frac{{1 - \delta }}{{1 + \delta }}} \right)}^2}}}} \arctan \left( {\sqrt {\frac{{{q^2} + \left( {1 - {q^2}} \right){{\left( {\frac{{1 - \delta }}{{1 + \delta }}} \right)}^2}}}{{1 - {{\left( {\frac{{1 - \delta }}{{1 + \delta }}} \right)}^2}}}} \frac{{{\text{sn}}\left( {\lambda x,q} \right)}}{{{\text{dn}}\left( {\lambda x,q} \right)}}} \right)} \right]} \right), \hfill \\ 
\end{gathered} 
\end{equation}
\end{widetext}

\subsection{$\alpha<0$}

For the negative value of $\alpha$ we need to solve Eq. (\ref{u07}) together with Eq. (\ref{polynomial}) corresponding to the case of $\alpha<0$. It is not necessary to repeat the solving procedure since one can change ${\lambda x}$ for ${i\lambda x}$ due to the presence the sign ``$-$'' before the square of the first derivative in Eq. (\ref{u07}). The imaginary argument transformation for the Jacobi elliptic functions can be applied to Eq. (\ref{psi_solution1}), (\ref{psi_solution2}) and (\ref{psi_solution3}) \cite{Abramowitz} and we obtain for real roots of Eq. (\ref{polynomial}) with the order ${r_1} \geqslant {r_2} \geqslant 0 \geqslant {r_3} \geqslant {r_4}$
\begin{equation}
\label{psi_solution1_1}
\psi  = {r_4} + \frac{{\left( {{r_1} - {r_4}} \right)\left( {{r_2} - {r_4}} \right)}}{{\left( {{r_2} - {r_4}} \right) - \left( {{r_1} - {r_2}} \right){\text{t}}{{\text{n}}^2}\left( {\lambda x,q'} \right)}},
\end{equation}
where ${\text{tn}}\left( {\lambda x,q'} \right) = \frac{{{\text{sn}}\left( {\lambda x,q'} \right)}}{{{\text{cn}}\left( {\lambda x,q'} \right)}}$ and $q' = \sqrt {1 - {q^2}}$. Eq. (\ref{psi_solution1_1}) admits the exact analytical integration
\begin{equation}
\label{u0_solution1_1}
u_0(x)=\frac{1}{\lambda }\left( {{r_1} - {r_2}} \right)\Pi \left( {{\text{sn}}\left( {\lambda x,q'} \right),\frac{{{r_1} - {r_4}}}{{{r_2} - {r_4}}},q'} \right) + {r_2}x,
\end{equation}

The same procedure for Eq. (\ref{psi_solution2}) yields 
\begin{equation}
\label{psi_solution2_1}
\psi  = {r_2} - \frac{{\left( {{r_2} - {r_3}} \right)\left( {{r_2} - {r_4}} \right)}}{{\left( {{r_2} - {r_4}} \right) + \left( {{r_3} - {r_4}} \right){\text{t}}{{\text{n}}^2}\left( {\lambda x,q} \right)}},
\end{equation}
and correspondingly 
\begin{equation}
\label{u0_solution2_1}
u_0(x)=\frac{1}{\lambda }\left( {{r_3} - {r_4}} \right)\Pi \left( {{\text{sn}}\left( {\lambda x,q'} \right),\frac{{{r_2} - {r_3}}}{{{r_2} - {r_4}}},q'} \right) + {r_4}x.
\end{equation}

\begin{figure}
\includegraphics[width=0.99\columnwidth]{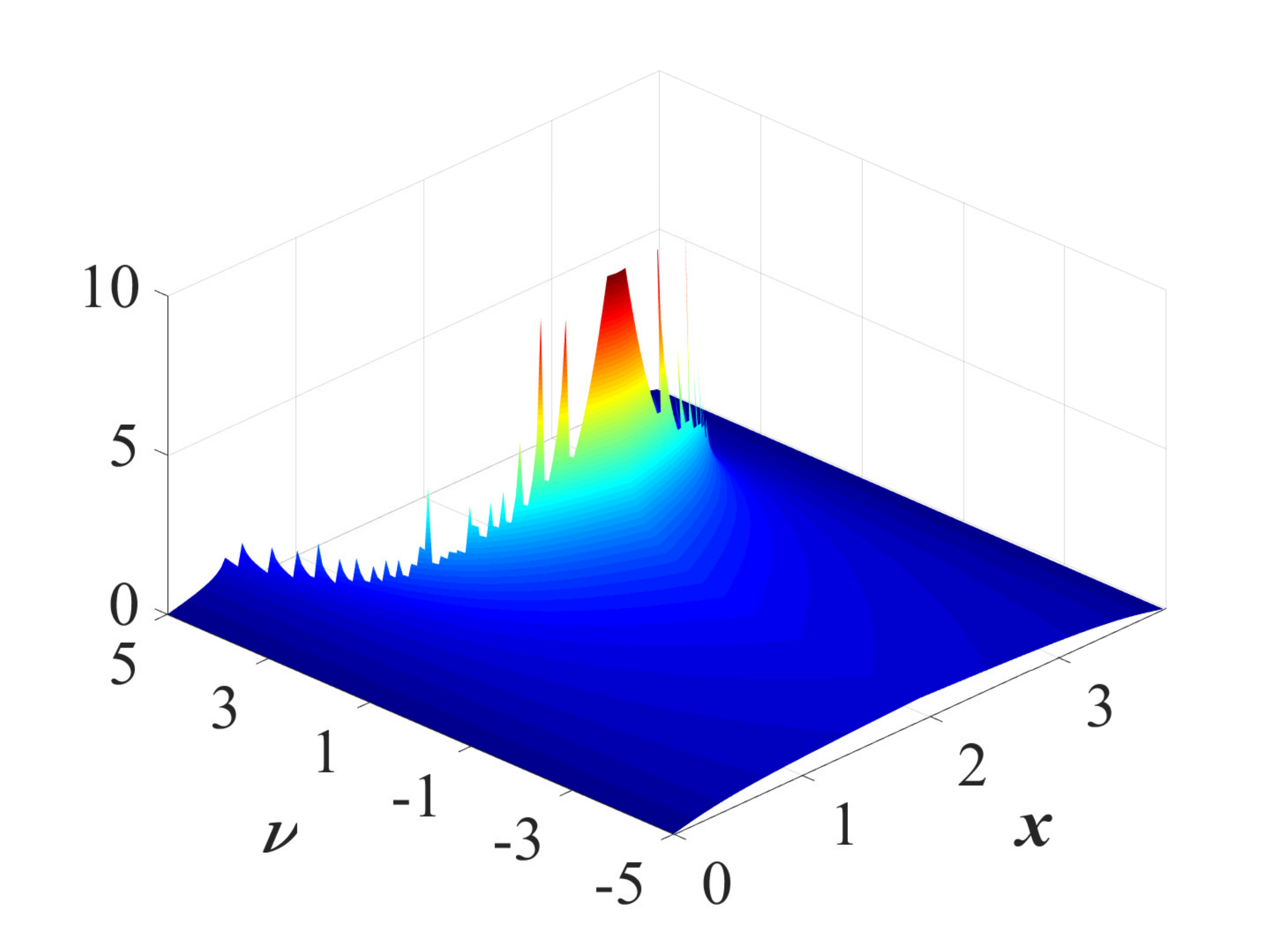}
\caption{The behavior of the function $\Pi \left( {{\text{sn}}\left( {x,m} \right),\nu ,m} \right)$ for $m=0.75$ vs the coordinate $x$ and the characteristic of the incomplete integral of the third kind $\nu$. The structure of the plot remains the same on the qualitative level for all $0<m<1$.}
\label{elliptic_integral_3}
\end{figure}

Finally, in the case of two real roots such as ${r_1} \geqslant {r_2}$ and two complex-conjugate roots $r_{3,4}=\rho_1 \pm i\rho_2$ in  Eq. (\ref{polynomial}) we have
\begin{equation}
\label{psi_solution3_1}
\psi  = {r_1} - {\frac{{\left( {{r_1} - {r_2}} \right)\left( {1 - {\text{nc}}\left( {\lambda x,q'} \right)} \right)}}{{1 + \delta  - \left( {1 - \delta } \right){\text{nc}}\left( {\lambda x,q'} \right)}}},
\end{equation}
where ${\text{nc}}\left( {\lambda x,q'} \right) = \frac{1}{{{\text{cn}}\left( {\lambda x,q'} \right)}}$. 

\begin{widetext}
\begin{equation}
\label{u0_solution3_1}
\begin{gathered}
  {u_0}\left( x \right) = \left( {{r_1} + \frac{{{r_1} - {r_2}}}{{1 + \delta }}} \right)x - \frac{{\left( {{r_1} - {r_2}} \right)}}{{2\lambda }}\frac{{1 - \delta }}{{1 + \delta }}\left[ {\Pi \left( {{\text{sn}}\left( {\lambda x,q'} \right),\frac{{{{\left( {1 + \delta } \right)}^2}}}{{4\delta }},q'} \right)} \right. \hfill \\
  \left. {\left. { + \frac{{1 + \delta }}{{1 - \delta }}\sqrt {\frac{{1 - {{\left( {\frac{{1 + \delta }}{{1 - \delta }}} \right)}^2}}}{{{{q'}^2} + \left( {1 - {{q'}^2}} \right){{\left( {\frac{{1 + \delta }}{{1 - \delta }}} \right)}^2}}}} \arctan \left( {\sqrt {\frac{{{{q'}^2} + \left( {1 - {{q'}^2}} \right){{\left( {\frac{{1 + \delta }}{{1 - \delta }}} \right)}^2}}}{{1 - {{\left( {\frac{{1 + \delta }}{{1 - \delta }}} \right)}^2}}}} \frac{{{\text{sn}}\left( {\lambda x,q'} \right)}}{{{\text{dn}}\left( {\lambda x,q'} \right)}}} \right)} \right]} \right) \hfill \\ 
\end{gathered}
\end{equation}
\end{widetext}

As one can see for all types of solutions we have similar structure that is characterized by the presence of the composite function $\Pi \left( {{\text{sn}}\left( {x,m} \right),\nu ,m} \right)$.  The selection procedure is based on the plot of this function in a dependence of the coordinate $x$ and the characteristic of the incomplete integral of the third kind $\nu$ (Fig. \ref{elliptic_integral_3}). Figure (\ref{elliptic_integral_3}) clearly demonstrates the dramatic enhancement with pronounced spikes occurs when $\nu>0$. Thus, among the plenty of solutions we need to choose those which have positive value of $\nu>0$.


\begin{thebibliography}{99}
\bibitem{note0}
Although we speculate about the hydrodynamic analogy to predict the phonon tsunami emergence, the wedge-shaped sample will be treated solely as a nonlinear elastic medium. In the following, therefore, the theoretical formalism will be based on the theory of elasticity. Such formal analogies, when used cautiously, allow to yield new occurrences. Examples of this are the liquid drop model of an atomic nucleus and the wave equation in the sting theory.
\bibitem{footnote1}
In the case of a thin plate, longitudinal modes obey a system of wave equations where components $u_x$ and $u_y$ of the displacement field are coupled through the mixed partial derivatives and with the different depending on Poisson’s ratio coefficients at the second order derivatives \cite{Landau}. Under certain assumptions this system of equations can be decomposed to separated wave equations for $u_x$ and $u_y$. This means that longitudinal modes are of the same nature as in the ``bulky'' 3D medium but with different velocities of the oscillations. Regardless of the type of imposed boundary conditions whether they are free (i.e., the first spatial derivatives of the displacement components are equal to zero) or they are clamped (the displacement components are equal to zero) in either case eigenfrequencies coincide.
\end{thebibliography}
\end{document}